\documentclass[]{emulateapj}
\usepackage{times, graphicx, setspace, subfigure, latexsym, amssymb, amsmath, fancyhdr, epsf, upgreek}
\usepackage[backref,breaklinks,colorlinks,citecolor=blue]{hyperref}
\usepackage{natbib}
\usepackage{multirow}
\usepackage{color}
\usepackage{rotating}
\usepackage{subfigure}
\usepackage{cprotect}
\usepackage{afterpage}
\usepackage{xspace}
\newcommand{\psrchive}{\textsc{psrchive}\xspace}
\newcommand{\pypulse}{\textsc{pypulse}\xspace}

\newcommand{\tempo}{\textsc{tempo}\xspace}
\newcommand{\temperr}{$\sigma_{\rm{S/N}}$\xspace}
\newcommand{\DISSerr}{$\sigma_{\rm{DISS}}$\xspace}
\newcommand{\jittererr}{$\sigma_{\rm{J}}$\xspace}
\newcommand{\residerr}{$\sigma_{\cal{R}}$\xspace}
\newcommand{\sigmac}{$\sigma^2_{C}$\xspace}
\newcommand{\scintt}{$\Delta t_{\rm{d}}$\xspace}
\newcommand{\scintnu}{$\Delta \nu_{\rm{d}}$\xspace}
\newcommand{\scinttau}{$\tau_{\rm{d}}$\xspace}
\newcommand{\RNum}[1]{\uppercase\expandafter{\romannumeral #1\relax}}

\newcommand{\biborder}[1]{}

\shorttitle{MSP Long Observation Analysis}
\shortauthors{B. J. Shapiro-Albert et al.}

\begin{document}

\title{Analysis of Multi-Hour Continuous Observations of Seven Millisecond Pulsars}


\author{B. J. Shapiro-Albert\altaffilmark{1,2}}

\author{M. A. McLaughlin\altaffilmark{1,2}}

\author{M. T. Lam\altaffilmark{3,1,2}}

\author{J. M. Cordes\altaffilmark{4}}

\author{J. K. Swiggum\altaffilmark{5}}

\altaffiltext{1}{Department of Physics and Astronomy, West Virginia University, Morgantown, WV 26506, USA}

\altaffiltext{2}{Center for Gravitational Waves and Cosmology, West Virginia University, Chestnut Ridge Research Building, Morgantown, WV 26506, USA}

\altaffiltext{3}{School of Physics and Astronomy, Rochester Institute of Technology, Rochester, NY 14623, USA}

\altaffiltext{4}{Cornell Center for Astrophysics and Planetary Science and Department of Astronomy, Cornell University, Ithaca, NY 14853, USA}

\altaffiltext{5}{Center for Gravitation, Cosmology and Astrophysics, Department of Physics, University of Wisconsin-Milwaukee, P.O. Box 413,
Milwaukee, WI 53201, USA}

\begin{abstract}
Precision pulsar timing can be used for a variety of astrophysical tests from the detection of gravitational waves to probing the properties of the interstellar medium (ISM). Here we present analyses of the noise contributions to pulsar timing residuals from continuous multi-hour observations of seven millisecond pulsars (MSPs). We present scintillation bandwidth measurements for all MSPs in the sample, some for the first time, and scintillation timescale measurements and lower limits for all MSPs for the first time. In addition, we present upper limits on the contribution of pulse phase jitter to the timing residual error for all MSPs. These long observations also allow us to constrain variations in dispersion measures (DMs) on hour-long timescales for several millisecond pulsars. We find that there are no apparent DM variations in any of the MSPs studied on these timescales as expected. In light of new radio telescopes such as the Canadian Hydrogen Intensity Mapping Experiment (CHIME), which will be able to time many pulsars for a short time each day, we search for differences in timing precisions from continuous TOAs and from equivalent length time-discontinuous TOAs. We find no differences in the precision for any MSP in our sample, as expected. We conclude that the TOA variations are consistent with the expected breakdown into template-fitting, jitter, and scintillation errors.
\end{abstract}


\hypersetup{linkcolor=blue}

\section{Introduction}

 Analyses of pulse times of arrival (TOAs) from pulsars, or pulsar timing, can be used to study many astrophysical phenomena. Not only can the objects themselves be studied, constraining their masses and equations of state \citep[e.g.,][]{Antoniadis2013,Stovall2018,Cromartie2019}, they can also be used as laboratories to probe extreme limits of general relativity \citep[e.g.,][]{Kramer2006, Archibald2018, Zhu2019}. In addition, groups like the North American Nanohertz Observatory for Gravitational Waves \citep[NANOGrav;][]{McLaughlin2013}, the European Pulsar timing array \citep[EPTA;][]{Kramer2013}, and the Parkes Pulsar Timing Array \citep[PPTA;][]{Hobbs2013} use pulsar timing arrays (PTAs) made up of millisecond pulsars (MSPs) to search for gravitational waves from supermassive black hole binaries \citep[e.g.][]{Shannon2013, Zhu2014, Lentati2015, Shannon2015, Arzoumanian2016, Babak2016, Verbiest2016, Arzoumanian2018_GW, Aggarwal2019}. In particular cases, continuous long observations such as those done by \cite{Dolch2014} can be used to produce single source gravitational wave limits \citep{Dolch2016}. Pulsar timing can also be used to study the properties of the interstellar medium (ISM) and how they change on timescales from hours to years \citep[e.g.,][]{Coles2015, Levin2016, Lam2016b, Jones2017}. 

For astrophysics that requires extremely precise pulsar timing, such as the detection of gravitational waves, every source of noise must be well modeled \citep[e.g.][]{Lam2018a}. As pulses propagate through the ISM, they are subjected to dispersion, scattering, and interstellar scintillation \citep{Rickett1977} which can each be a source of noise \citep[][]{Shannon2012, Lam2016a, Lam2019} in the data. Interstellar scintillation is not correlated between observing epochs and thus will present as white (uncorrelated in time) noise.
Changes in the pulse shape with frequency \citep[][]{Kramer1998, Pennucci2014} along with interstellar scintillation are another source of white noise. Variations in pulse phases and amplitudes, or pulse jitter, \citep[][]{Shannon2010,Shannon2012,Lam2019} also appear as white noise. Dispersion and scattering are sources of red (correlated in time) noise, and stochastic variations in the pulse spin rate also manifest as red noise \citep[][]{Cordes1986, Cordes1998, Lam2016a}.

The first source of TOA error we consider is from additive noise that causes template-fitting errors. Time-averaged pulse templates are cross-correlated with the observed pulses to determine the TOA. For an observed pulse averaged over some number of pulses, $N_{p}$, the precision of the TOA, or template-fitting error, goes as $1/\sqrt{N_{p}}$ \citep[][]{Taylor1992, Dolch2014}. However, if the pulse shape varies or has a low signal-to-noise ratio (S/N), then there may be additional errors above what is expected due to additive noise alone \citep{Arzoumanian2015}.

Pulse jitter is a second source of TOA error and occurs due to motions of coherent emission regions in pulsar magnetospheres \citep{Cordes2010}. While average pulse profiles are highly stable in time \citep[e.g.][]{Brook2018}, for single pulses the phase and amplitude can be highly variable. As single pulses from MSPs are generally very weak, the pulse jitter is often difficult to measure directly from single pulses.

Diffractive interstellar scintillation (DISS), in combination with the fact that the pulse shape changes with observing frequency, cause a third source of TOA error \citep{Liu2014, Pennucci2014}.
Diffractive scintillation due to the ISM causes constructive interference between the pulse ray paths which leads to increases in pulse intensity (and thus S/N) over particular frequencies, or ``scintles" \citep{Cordes1986, Cordes1998}. These variations in the pulse S/N with the pulse shape changes in frequency may change the shape of the frequency-averaged pulse profile which would cause template-fitting errors.

DISS can also cause the pulses to be broadened, inducing additional time-variable delays \citep{Hemberger2008}. However, for a turbulent (Kolmogorov) medium the average frequency scale of the interference, or scintillation bandwidth (\scintnu), and the average duration, or scintillation timescale (\scintt), can be measured. Given a set of assumptions about the ISM, such as a homogeneous turbulent medium  and spatial scale of the diffraction pattern larger than the observing baseline, the scintillation bandwidth can inform the magnitude of scattering and constrain the impact on pulsar timing \citep{Cordes1986, Levin2016, Lam2016b, Lentati2017}. However, as there will be a finite number of scintles on the frequency-time plane, our ability to accurately measure \scintnu and \scintt is limited by the number of observed scintles \citep{Cordes1990}. This will also cause the scatter broadening function to be stochastic in what is known as the ``finite scintle effect."

 Finally,  time delays due to dispersion are $\propto$ DM $\times$ $\nu^{-2}$, where the dispersion measure (DM) is the integrated column density of free electrons along the line of sight and $\nu$ is the frequency of the radio emission.  Since the ISM is a turbulent medium the DM can change with time and, if unmodeled, it can be a source of chromatic red noise \citep{Keith2013, Jones2017}.

Large scale studies of pulse jitter, scintillation parameters, and subsequent timing errors have been done by \cite{Lam2016a} and \cite{Levin2016} respectively on the NANOGrav 9-yr data set \citep{Arzoumanian2015} and an updated analysis of the pulse jitter has also been completed for the NANOGrav 12.5-yr data set \citep{Lam2019}. However, the previous scintillation parameter measurements were limited by the typical $\sim$25~minute observation lengths. While it is possible to characterize \scintnu on this timescale, \scintt is almost always $\gtrsim 30$~minutes for these pulsars at the radio frequencies observed. This limits how accurately the DISS effects on pulsar timing can be estimated.

Additionally, \cite{Jones2017} measured the DM variations and evolution for 37 MSPs in the NANOGrav 9-yr data set to mitigate the chromatic red noise and found that the DM varies on timescales of days to years. However, due to the cadence of their observations, they were unable to probe variations on timescales shorter than $\sim14$~days. Studies of DM variations on shorter timescales \citep[e.g.][]{Hankins2016} can additionally inform us about the ISM along a particular line of sight and its effects on precision pulsar timing. However, either higher cadence observations or longer observations are required to look for DM variations on these timescales.

Here we present our analyses of eight continuous multi-hour observations of seven MSPs, all part of the NANOGrav PTA. These observations allow us to study the scintillation parameters, pulse jitter, and DM variations on $\sim$hour-long timescales (along each particular line of sight), similar to analysis done by \cite{Dolch2014} on a 24~hour multi-band continuous observation of PSR~J1713$+$0747.

The standard NANOGrav timing procedure is to observe each MSP for typically $\sim$25~minutes every few weeks \citep{Arzoumanian2015, Arzoumanian2018}. Conversely, telescopes like the Canadian Hydrogen Intensity Mapping Experiment (CHIME), will make daily observations of multiple MSPs more common \citep{Ng2018}. However, each pulsar may only be visible to CHIME for $\sim5$~minutes daily. The length of our observations present an opportunity to test the timing precision of a contiguous observation versus the same amount of time but split into several short observations (non-contiguous), where the pulse S/N will be lower and the template-fitting error may increase beyond the expected $\propto 1/\sqrt{N_p}$. This is useful in considering how adding CHIME pulsar timing data to NANOGrav will affect the timing precision.

We describe the observations and the basic data reduction pipeline in \S \ref{observations}. Our methods for analyzing the pulse jitter, scintillation parameters, DM variations on short timescales, and timing precision of non-contiguous observations are described in \S \ref{methods}. We present and discuss the results of our scintillation and ISM analysis in \S \ref{scintparamsss}. The pulse jitter analyses are presented and discussed in \S \ref{allthejitter}. The results of our DM variations analysis are discussed in \S \ref{dmvars}. Finally, we present the results of testing our timing precision with non-contiguous TOAs in time in \S \ref{ResidChunkResults}. We offer concluding remarks in \S \ref{conclusion}.

\section{Observations} \label{observations}

\subsection{Observational Data}
We observed seven NANOGrav MSPs (PSRs~J0023$+$0923, J0340$+$4130, J0613$-$0200, J0645$+$5158, J1614$-$2230, J1832$-$0836, and J1909$-$3744) for between $\sim$1.5 and $\sim$6~hours each between MJDs~56724 (2014 March 8) and 56842 (2014 July 4). All observations were taken with the Robert C. Byrd Green Bank Telescope (GBT) at the Green Bank Observatory. Each observation had a center frequency of 1500~MHz and a bandwidth of 800~MHz with 1.5625~MHz frequency resolution. The raw profiles were folded in $\sim$15~s integrations in real time and coherently dedispersed at the DM listed in Table \ref{mspinfo} by the GUPPI backend \citep{DuPlain2008}. Each pulse profile was divided into 2048 phase bins and recorded with two polarizations.

Over the course of the observations, instrumental difficulties caused parts of our frequency band to be lost in two of our observations, the first observation of PSR~J0645$+$5158, and the observation of PSR~J1909$-$3744. The PSR~J0645$+$5158 observation was restarted and recalibrated, resulting in a $\sim$1~hour gap between the two segments; the length of the sections are 1.5 and 2.5 hours, and were analyzed separately. Our observation of PSR~J1909$-$3744 experienced similar data acquisition difficulties throughout the observation, resulting in the number of frequency channels being recorded dropping from 512 to 448, to 256, and finally to 192. We did not analyze the 448 channel section of the observation in this work due to errors in the header of the data file. The other three sections of the observation were analyzed separately. The length of each observation can be found in Table \ref{mspinfo}.

\begin{deluxetable*}{lcccccccc}[h!]
\tablecaption{Pulsar Parameters \label{mspinfo}}
\tablecolumns{8}
\tablehead{
\colhead{PSR Observation} & \colhead{R.A. (J2000)} & \colhead{Dec (J2000)} & \colhead{$P$} & \colhead{$\dot{P}$} & \colhead{DM} & \colhead{MJD} & \colhead{Observation Length} & \colhead{Bandwidth}\\
\colhead{} & \colhead{(hms)} &
\colhead{($^{\circ}$ $' "$) } & \colhead{(ms)} & \colhead{$10^{-20}$ (s/s)} & \colhead{(pc cm$^{-3}$)} & \colhead{} & \colhead{(hrs)} & \colhead{(MHz)}
}
\startdata
J0023$+$0923 & 00:23:16.87 & $+$09:23:23.86 & 3.05 & $1.14$ & 14.32 & 56732 & 2.85 & 800 \\
J0340$+$4130 & 03:40:23.28 & $+$41:30:45.29 & 3.30 & $0.70$ & 49.59 & 56724 & 4.00 & 800\\
J0613$-$0200 & 06:13:43.97 & $-$02:00:47.23 & 3.06 & $0.96$ & 38.78 & 56733 & 5.76 & 800\\
J0645$+$5158 (1) & 06:45:59.08 & $+$51:58:14.91 & 8.85 & $0.49$ & 18.25 & 56726 & 1.62 & 800\\
J0645$+$5158 (2) & & &  & &  & 56726 & 2.46 & 800\\
J0645$+$5158 (3) & & & & & & 56736 & 1.66 & 800\\
J1614$-$2230 & 16:14:36.50 & $-$22:30:31.27 & 3.15 & $0.96$ & 34.49 & 56842 & 3.37 & 800\\
J1832$-$0836 & 18:32:27.59 & $-$08:36:55.01 & 2.72 & $0.83$ & 28.19 & 56841 & 5.61 & 800\\
J1909$-$3744 (512 channels) & 19:09:47.43 & $-$37:44:14.46 & 2.95 & $1.40$ & 10.39 & 56758 & 1.35 & 800\\
J1909$-$3744 (256 channels) & & & & & & 56758 & 1.80 & 400\\
J1909$-$3744 (192 channels) & & & & & & 56758 & 2.98 & 300
\enddata
\tablecomments{General pulsar parameters of the MSPs observed in our data set. All values are from the NANOGrav 11-yr data release \citep{Arzoumanian2018}. For PSR~J0645$+$5158, the first observation has a $\sim$1~hr gap between two halves of the observation so they are labeled (1) and (2). A second observation of the same MSP 10~days later is denoted (3). Each segment/observation is analyzed separately. For PSR~J1909$-$3744, technical issues caused frequency channels to be dropped throughout the observation. The corresponding bandwidth of each observation is reported above. Each observation is denoted by the number of frequency channels recorded in that segment and analyzed separately.}
\end{deluxetable*}

\subsection{Data Reduction}

The flux and polarization calibration procedures as well as initial radio frequency interference (RFI) mitigation techniques closely follow those of \cite{Arzoumanian2018}. Our data reduction and analysis makes use of both the \psrchive\footnote{\url{http://psrchive.sourceforge.net/index.shtml}} software package \citep{Hotan2004, VanStraten2012} and the \textsc{python} software package \pypulse\footnote{\url{https://github.com/mtlam/PyPulse}} \cite{LamPyPulse}.

The polarization calibration observation was performed by injecting a broadband noise signal into both polarizations at the telescope before beginning the observation, and recording it with the GUPPI backend. We then calibrated both the phase angle between the polarizations and the differential gain with the noise signal with \psrchive. Flux calibration was obtained from NANOGrav observations of the radio source B1442$+$101 closest to the date of each MSP observations taken with the same receiver. Full intensity profiles were obtained by summing the two polarizations of each profile together.

While a  polarization calibration scan was done at the beginning of each of our observations and polarization cross-products
were recorded, we only use total intensity measurements, obtained by summing the calibrated
signals from pairs of orthogonal polarizations.  This lack of calibration for feed coupling  could produce time-variant profiles and TOAs at different parallactic angles \citep{Liu2011}. However, this is the same procedure as used for the NANOGrav 11-year dataset \citep{Arzoumanian2018} and  there is no evidence of flux or profile variations due to incorrect polarization calibrations in those observations \citep{Brook2018}. We therefore expect excess noise to be minimal.

RFI mitigation was performed using the \psrchive software. We first removed frequency channels known to be contaminated by RFI as denoted in the NANOGrav data reduction pipeline \citep{Demorest2013}. We then removed frequency channels and integrations where the off-pulse variance within a 20-channel/integration wide window was more than four times the median channel variance. Much of our data was heavily contaminated by RFI, so this off-pulse variance mitigation method was then rerun with a threshold of three times the median channel variance. Data were then checked manually to verify that RFI mitigation was successful, and any remaining RFI was manually removed.

For each particular analysis of the data, each observation was integrated in time and/or frequency using either the \psrchive or \pypulse packages to build up the S/N and/or minimize computation time. The various subsections in \S \ref{methods} detail the subsequent data processing for each analysis.

\section{Methods} \label{methods}

Here we lay out the methods for all analyses performed on our long observations. We first describe how we determined the total rms of our timing residuals and estimated the individual noise contributions. We discuss multiple ways to detect pulse jitter as in \cite{Shannon2012, Shannon2014, Lam2016a, Lam2019}. We then describe how the scintillation parameters, \scintnu and \scintt, were determined.

We also lay out methods for calculating pulsar secondary spectra to study the ISM along the line of sight as in \cite{Stinebring2000, Stinebring2019}. We then detail our methods for constraining short timescale DM variations. Finally we describe how these long observations were used to assess the accuracy of non-contiguous pulsar TOAs when compared to TOAs generated from a contiguous time series.

\subsection{White Noise in Pulsar Timing Residuals} \label{whitenoise}

White noise in pulsar timing residuals on short timescales is composed of three components: template-fitting errors, \temperr, which are dependent on the pulse S/N, DISS variations, \DISSerr, and errors due to intrinsic pulse jitter, \jittererr. We do not address errors due to calibration or residual RFI. The total white noise error contribution to our residuals, or their rms, \residerr, can be characterized by
\begin{equation} \label{whitenoiseerrs}
    \sigma^{2}_{\cal{R}} = \sigma^2_{\rm{S/N}} + \sigma^2_{\rm{J}} + \sigma^2_{\rm{DISS}}.
\end{equation}
For most MSPs \temperr $>$ \jittererr $\gg$ \DISSerr. However, in the high S/N regime, we may observe \jittererr $\gtrsim$ \temperr.

We used the pulsar timing packages \psrchive and \pypulse package to generate residuals and calculate \residerr. For the duration of this work we followed the methods of \cite{Lam2016a} to generate ``short-term" timing residuals, ${\cal{R}}$($\nu,t$), for each observation.
We used the NANOGrav 11-yr timing parameters to fold our data. We assumed that these timing models were sufficiently accurate for our data sets so no model parameters were fit for. 

However, since each epoch was analyzed separately, we determined only the pulse phase within an observation, or ``initial timing residuals", $\delta t$($\nu,t$), and did not use the NANOGrav 11-yr timing parameters to determine the timing residuals. This method assumes that after using this timing model to fold the data we will be left with a polynomial expansion of pulse phase and spin period representative of the Earth-pulsar line of sight at the given epoch. These short-term residuals were calculated using the Fourier-domain estimation algorithm of \cite{Taylor1992}. Following \cite{Lam2016a}, we then calculated the timing residuals, ${\cal{R}}$($\nu,t$), by fitting a polynomial over $\delta t$($\nu,t$) that included a constant offset for TOAs from each frequency channel and parabolic term common to all TOAs in time:
\begin{equation} \label{intrinsicTOAs}
    \delta t (\nu,t) = K (\nu) + at + bt^{2} + n(\nu,t),~{\rm{and}}
\end{equation}
\begin{equation} \label{shorttermTOAs}
    {\cal{R}}(\nu,t) \equiv \hat{n}(\nu,t) = \delta t(\nu,t) - \left [ \hat{K} (\nu) + \hat{a}t + \hat{b}t^{2} \right ].
\end{equation}
Here $a$ and $b$ are frequency-independent coefficients, $n$($\nu,t$) is additive white noise in frequency and time, including all components in Eq. \ref{whitenoiseerrs}, and $K(\nu)$ is a constant offset in frequency that accounts for frequency dependent variations such as profile evolution and scattering. The polynomial fit provides a simple way to remove deviations from the ``true'' timing model by use of the initial timing residuals. All values with carats are estimated quantities. Therefore Eqs. \ref{intrinsicTOAs} and \ref{shorttermTOAs} denote ${\cal{R}}$($\nu,t$) as the estimated additive noise where frequency dependence between sub-bands has been subtracted off.

Determining \residerr as a function of integration time allowed us to both extrapolate the expected \residerr for a single pulse and check that \residerr $\propto 1/\sqrt{N}$, where now $N$ is the total integration time and is proportional to $N_p$. In particular we have used integration times of 15 and 30 seconds as well as 1, 2, 4, 8, 16, and 32 minutes all with 12.5~MHz width per channel as used in the NANOGrav timing analysis \citep{Arzoumanian2018} resulting in 64 residuals per integration. We note again that the quadratic in Eq. \ref{shorttermTOAs} is not fit per channel and therefore there still remains a significant amount of white noise over the measurements from each of the 64 channels even when the mean of those is subtracted by estimating $\hat{K}$. Additionally, for longer integration times, the last integration was dropped if the length was not comparable to our desired time integration length. For our shortest observations with 32 minute integration times, we will have at minimum 128 residuals, so any variance that is absorbed using Eqs. \ref{intrinsicTOAs} and \ref{shorttermTOAs} will be very small \citep{Lam2016a}.

Assuming that all pulses emitted by the pulsar are statically independent we were additionally able to test how well \residerr values follow the $1/\sqrt{N}$ relationship by fitting not only for the value of \residerr at a integration time of a single pulse, but also for the slope as $1/N^{\alpha}$. Any deviations from a slope of 0.5 would show that the pulses are not statistically independent \citep{Helfand1975, Rankin1995}.

Standard methods of pulsar timing assume the observed pulse is a scaled and shifted version of the pulse profile with added noise. For obtaining pulse TOAs using matched filtering, this assumption yields the minimum TOA error. Again following the formalism of \cite{Lam2016a}, for a pulse with some effective width, $W_{\rm{eff}}$, and $N_{\phi}$ phase bins, we write the template-fitting error \citep{Cordes2010}
\begin{equation} \label{tempfittingerrs}
    \sigma_{\rm{S/N}} = \frac{W_{\rm{eff}}}{S \sqrt{N_{\phi}}},
\end{equation}
where $S$ is the S/N of the pulse taken as the peak to off-pulse rms ratio. Since the pulse S/N is easily measured in this way, \temperr is easily calculated. 

Similarly, $W_{\rm{eff}}$ of the pulse is dependent on both the pulse period, $P$, and the pulse template shape $U(\phi)$ as \citep{Downs1983}
\begin{equation} \label{WeffEq}
    W_{\rm{eff}} = \frac{P}{N_{\phi}^{1/2} [\sum_{i=1}^{N_{\phi}-1} [U(\phi_{i}) - U(\phi_{i-1})]^{2} ]^{1/2}}.
\end{equation}
We note that $W_{\rm{eff}}$ was calculated separately for each pulsar and can be determined separately for each frequency or backend. As all of our observations were taken at a central frequency of 1500~MHz with the GUPPI backend, we used those parameters to determine $W_{\rm{eff}}$ for all of the MSPs in our data set. However, both pulse jitter and scintillation can dynamically change the pulse profile which requires additional errors to be considered \citep{Cordes1985}.

\subsection{Scintillation Parameters and \DISSerr} \label{scintparams}

 As the ISM is dynamic, the frequency-dependent diffraction of the pulses due to the ISM will change as a function of time. This diffraction varies the path length of the pulses with time causing the pulse broadening function to change the pulse shape which is the source of \DISSerr \citep{Cordes1990, Lam2016a}. While it is difficult to determine exactly how much the ISM is broadening the pulse due to covariances with frequency-dependent intrinsic pulse shape variations, the resulting scintillation pattern, or dynamic spectrum, has a characteristic timescale, \scintt, and frequency scale, \scintnu. The scattering timescale, $\tau_{\rm{d}}$, is related to \scintnu by 
\begin{equation} \label{scateq}
    \tau_{\rm{d}} = \frac{C_{1}}{2 \pi \Delta \nu_{\rm{d}}},
\end{equation}
where $C_{1}$ is a coefficient ranging from 0.6--1.5 depending on the geometry and spectrum of the electron density of the ISM \citep{Lambert1999}. Here we set $C_{1} = 1$ as done in \cite{Levin2016}. If \scintnu can be measured, the TOA error due to pulse scattering can be estimated directly as
 \begin{equation} \label{DISSeq}
    \sigma_{\rm{DISS}} \approx \frac{\tau_{\rm{d}}}{\sqrt{n_{\rm{ISS}}}},
\end{equation}
where $n_{\rm{ISS}}$ is the number of scintles observed \citep{Cordes2010}. This determines our ability to accurately measure \scintnu, \scintt, and thus \DISSerr.

For Eq. \ref{DISSeq} to be true, $n_{\rm{ISS}}$ must be large. If \scintt and \scintnu, can be measured, $n_{\rm{ISS}}$ for a single observation of length $T$ and total bandwidth $B$ can be estimated from
\begin{equation} \label{nscintles}
    n_{\rm{ISS}} \approx \left ( 1 + \eta_{t} \frac{T}{\Delta t_{\rm{d}}} \right ) \left ( 1 + \eta_{\nu} \frac{B}{\Delta \nu_{\rm{d}}} \right ).
\end{equation}
Here $\eta_{t}$ and $\eta_{\nu}$ are filling factors in the range 0.1--0.3 \citep{Cordes2010}, which were both set to 0.2 as in \citet{Levin2016}.

For many NANOGrav observations, \scintt$> T$ and is not measurable, so $(1 + \eta_{t} (T / \Delta t_{\rm{d}})) \approx 1$. However, if \scintt is more accurately measured, $n_{\rm{ISS}}$ and thus \DISSerr, can be more accurately estimated. After determining \DISSerr Eq. \ref{whitenoiseerrs} can then be solved for \jittererr.

\begin{figure*}[p]
    \centering
    \includegraphics[width=\textwidth,height=\textheight,keepaspectratio]{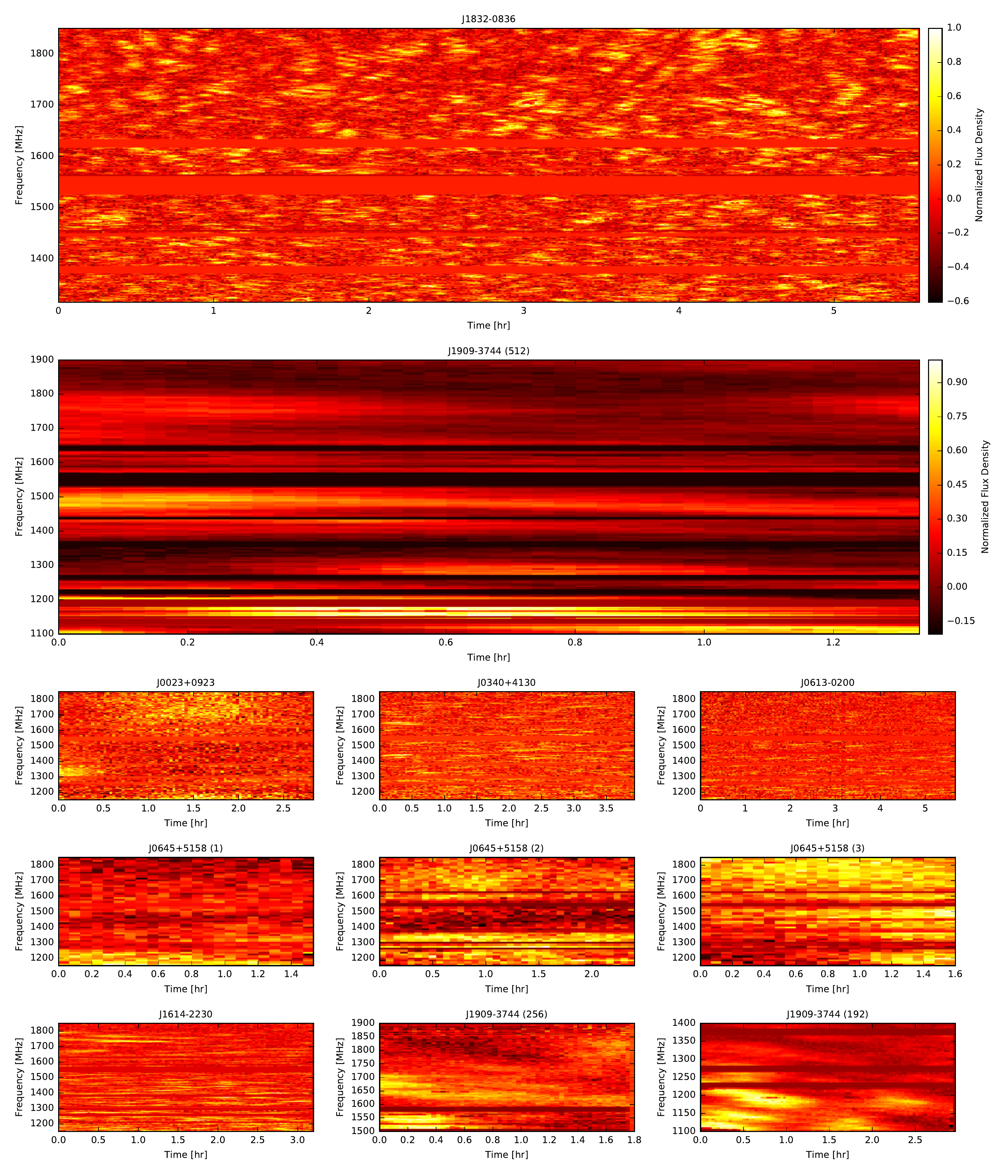}
    \caption{Unstretched dynamic spectrum for all MSPs analyzed in this work. Bright (yellow) patches show scintles. RFI zapped channels had their values replaced with the mean value of the dynamic spectrum. All values are reported in terms of normalized flux density. The two large dynamic spectra (PSRs~J1832$-$0836 and J1909$-$3744) are shown to contrast narrow, short scintles with wide, long scintles.}
    \label{DynSpec_Panels}
\end{figure*}

\subsubsection{Measuring Scintillation Parameters} \label{scintparamsmethods}

As our observations ranged from $\sim$1.5--6 hours in length and were taken, in most cases, over an 800~MHz bandwidth, we expected $n_{\rm{ISS}}$ would be large based on previous measurements from \cite{Levin2016} and estimates from the NE2001 electron density model \citep{NE2001}. We therefore attempted to measure \scintt and \scintnu independently for each MSP. We did this using two methods.

In order to fit the scintillation parameters, after our initial data reduction we created a 2-D dynamic spectrum for each MSP using the \pypulse \textsc{python} package. The dynamic spectrum shows how the intensity of the pulsar emission varies as a function of both time, $t$, and frequency, $\nu$. In \pypulse this is performed by first subtracting baseline variations, and then taking a template profile, here the NANOGrav 11-yr templates, and using the template-matching procedure of \cite{Taylor1992} to calculate the peak amplitude, or intensity, of the profile in each time-frequency bin in mJy. These intensity values are used as the dynamic spectrum values. Any frequency channels that were zapped due to RFI were replaced with the mean power value of the full dynamic spectrum. 

While this peak-amplitude method differs from \cite{Levin2016}, which subtracts the off-pulse flux from the on-pulse flux and divides by the mean off-pulse value, we still obtain robust dynamic spectra. Our data were polarization and flux calibrated and the baseline is subtracted so we can be confident that no variations in the baseline are included in our peak-amplitude dynamic spectra values. The pulse profiles of all MSPs in this work are sharply peaked, so the peak-amplitude returned by the template-matching procedure will be a robust proxy for the total on-pulse flux. Thus any variations in the peak-amplitude values should therefore be due to scintillation.

For some pulsars it is obvious that \scintt and/or \scintnu are much larger than the initial integration time of 15~seconds or frequency channel size of 1.5625~MHz. In these cases, a coarser resolution in time and/or frequency was used to build up pulse S/N, and we integrated in time or frequency such that there were at least $\sim$5 integrations/frequency channels spanning each scintle. The resolution of the dynamic spectrum for each MSP can be seen in Figure \ref{DynSpec_Panels}.

To measure the characteristic scales for both \scintt and \scintnu over our band, we had to take into account the frequency dependence of both parameters. The frequency dependencies of these parameters differ however, with \scintt $\propto \nu^{1.2}$ \citep{Rickett1977} and \scintnu $\propto \nu^{4.4}$ \citep{Cordes1986}. With this relation, \scintt will change by less than a factor of two over the bandwidth. In addition, \scintt is often of order the length of the observation, whereas \scintnu is usually much smaller than our bandwidth, so we expect fewer scintles in time than in frequency.

However, the dependence of \scintnu on frequency is much steeper. To account for this, we adopted the same ``stretching" method used by \cite{Levin2016} and stretched the dynamic spectrum to a reference frequency of 1500~MHz, assuming the $\nu^{4.4}$ frequency dependence. 

\begin{figure}
    \centering
    \includegraphics[width = 9cm]{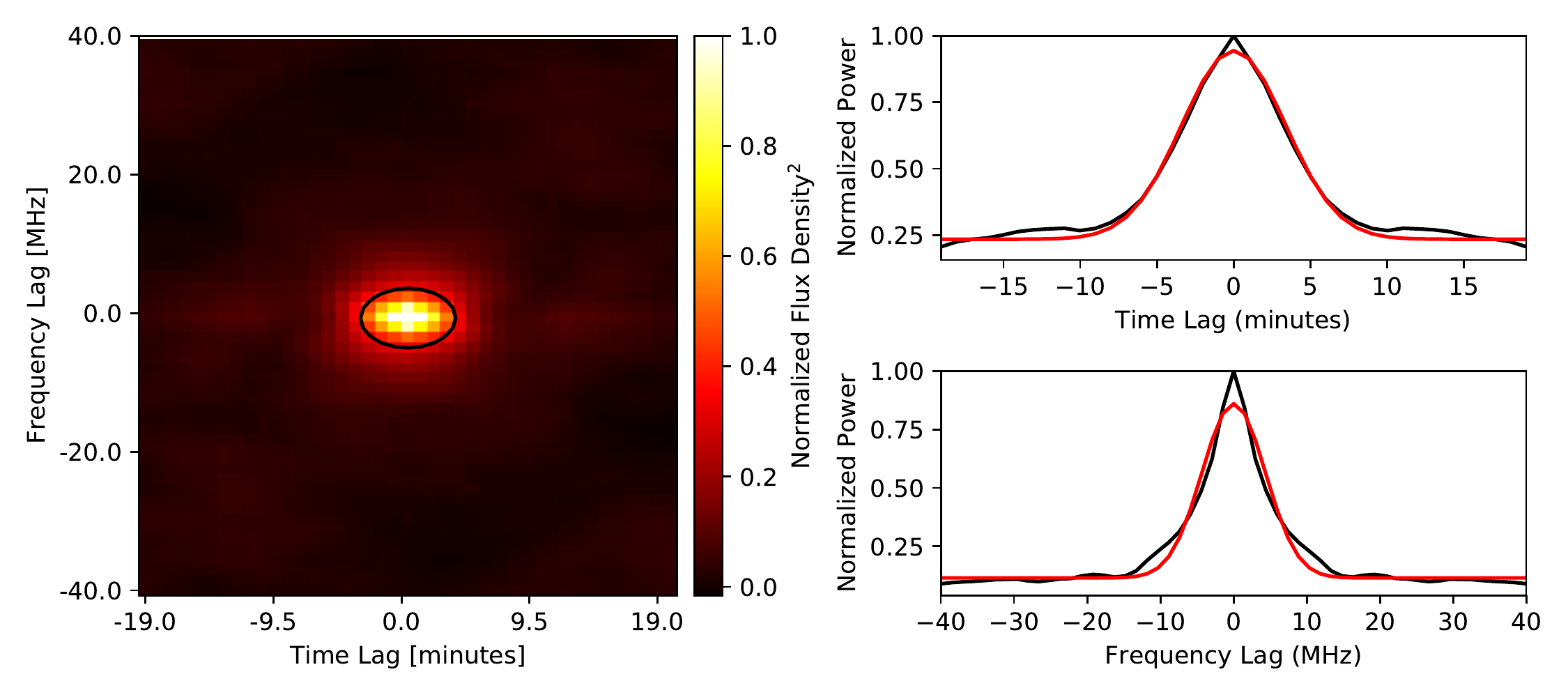}
    \caption{Left: 2-D Autocorrelation function (ACF) of the dynamic spectrum for PSR~J1832$-$0836 after stretching. The black contour shows the 2-D Gaussian fit of the 2-D ACF. Upper Right: 2-D ACF summed along the frequency axis with the central noise spike removed. The red line is the Gaussian used to obtain the scintillation timescale. Lower Right: Same as above but summed along the time axis with the Gaussian used to describe the scintillation bandwidth.}
    \label{ACFs_example}
\end{figure}

 We then computed the 2-D autocorrelation function (ACF) of each dynamic spectrum. An example can be seen in the left panel of Figure \ref{ACFs_example}. The first method used to estimate the scintillation parameters is similar to that of \cite{Levin2016}. We summed the 2-D ACF over a subsection of the time axis for \scintnu, or over a subsection of the frequency axis for \scintt, such that the central power region (in either time or frequency) was summed over without adding in noise. 
A Gaussian, centered at zero-lag (in either time or frequency), was then fit to each resulting 1-D ACF, as shown in the right panels of Figure \ref{ACFs_example} for example.

Often there is a noise spike centered at zero-lag in time and/or frequency which can bias the Gaussian fit. In order to minimize the effect of this spike, the value of the 1-D ACF at zero-lag was replaced with the average of the two points to either side of it. The value of \scintnu was taken to be the half-width at half max of the resulting Gaussian fit over the 1-D ACF vs time lag, and \scintt was the half-width at $e^{-1}$  of the Gaussian fit over the 1-D ACF vs frequency lag \citep{Cordes1986}.

The second method we used utilizes \pypulse to fit a 2-D Gaussian to the 2-D ACF. The central noise spike described above was again replaced with the average of these two points. We used just the central subsection of the 2-D ACF such that the full central power region was included with as little noise as possible. An example of the 2-D fit is shown by the black contour in the left panel of Figure \ref{ACFs_example}.

Since all of our observations are over an hour in length, if we were able to resolve \scintt, the 2-D fitting allowed us to better constrain and inform the fit values for \scintnu and \scintt. Additionally, the long observation times allowed us to fit for the drift rate in the scintles to constrain refractive variations through the ISM as \citep{Hewish1980,cpl1986}
\begin{equation}
\frac{d\nu}{dt} = \frac{\nu V_\perp}{2 d_s \theta_r},
\label{eq:driftrate}
\end{equation}
where $V_\perp$ is the pulsar's velocity perpendicular to the line of sight, $d_s$ is the distance between the Earth and a thin screen of material (in terms of the total distance $D$ and the pulsar-screen distance $D_s$, $d_s = D - D_s$), and $\theta_r$ is the component of the refractive angle along the direction of the pulsar's motion. This drift rate is fit for by a rotation of the 2-D Gaussian fitted to the 2-D ACF, in the frequency and time lags plane.

\subsection{Secondary Spectra}

The dynamic spectrum can also be used to study deviations from the typical thin screen model of scattering \citep{Scheuer1968}. The 2-D Fourier transform of the dynamic spectrum, or secondary spectrum, may show scitillation ``arcs" which can be used to study the structure of a scattering screen between us and the pulsar \citep{Stinebring2001,Cordes2006}. The shape, thickness, and number of arcs can be used to infer the location of the screen between us and pulsar, the thickness of the screen on $\sim$AU scales, and the number of screens causing the scattering \citep[][]{Stinebring2001,Stinebring2019}. 

Only two MSPs have already been seen to exhibit scintillation arcs, PSRs~J0437$-$4715 \citep{Bhat2016, Reardon2018} and B1937$+$21 \citep{Walker2013}, so these observations present an opportunity to search for scintillation arcs in these MSPs for the first time. Secondary spectra are generated with \pypulse by first taking a 2-D fourier transform of the dynamic spectrum, and then taking the square of the absolute magnitude of the transformed spectra \citep[][]{Stinebring2001,Reardon2019}. As the power of the secondary spectra is log-normally distributed, we took the log of the secondary spectra to better search for scintillation arcs.

\subsection{Maximum Likelihood Estimates of Pulse Jitter} \label{funnelplots}

Instead of directly solving Eq. \ref{whitenoiseerrs} using values obtained from Eqs. \ref{tempfittingerrs} and \ref{DISSeq} to determine the pulse jitter, we can also follow the methods of \cite{Lam2016a}. To estimate \jittererr from the timing residuals, we can rewrite Eq. \ref{whitenoiseerrs} as a function of the pulse S/N \citep{Lam2016a},
\begin{equation}
    \sigma^{2}_{\cal{R}}(S) = \sigma^{2}_{\rm{S/N}}(S) + \sigma^2_{\rm{DISS}} + \sigma^{2}_{\rm{J}} = \left (\frac{W_{\rm{eff}}}{S \sqrt{N_{\phi}}}  \right )^{2} + \sigma^2_{C},
\end{equation}
 where $\sigma^2_{C} = \sigma^2_{\rm{DISS}} + \sigma^{2}_{\rm{J}}$ is the rms that is ``constant'' in $S$ and $\sigma^{2}_{\cal{R}}(S)$ is calculated from the rms of ${\cal{R}}$($\nu,t$). If we can measure \sigmac, we can estimate \jittererr using \DISSerr obtained from \scintt and \scintnu. 
 
 To estimate \sigmac we performed the maximum likelihood analysis over the residuals detailed in \cite{Lam2016a}. This was done by assuming that, for a given S/N, our residuals will follow a Gaussian distribution described by
 \begin{equation}
     f_{{\cal{R}} \mid S}({\cal{R}} \mid S, \sigma_{C}) = \frac{1}{\sqrt{2 \pi \sigma^{2}_{\cal{R}}}} e^{-{\cal{R}}^{2}/(2\sigma^{2}_{\cal{R}})}.
 \end{equation}
From this we could determine the PDF of \sigmac from our residuals and predetermination of \temperr. This PDF was evaluated with \pypulse to determine the value of \sigmac as described in \cite{Lam2016a}.
An F-test with a significance value of 0.05 ($2\sigma$) was used to determine whether or not the determined value of \sigmac was significant. The 95\% upper limit on \sigmac is reported if it was not.

While \cite{Lam2016a} used a S/N cutoff of $\sim$3 to filter out low-significance noise from the data, we did not include a low S/N cutoff as many of our MSPs have few, if any, TOAs above this threshold. To mitigate noise in our data due to our lack of a low S/N cutoff, outlier residuals were removed via sigma stripping with a $3\sigma$ threshold where $\sigma$ is determined from the overall distribution of \residerr.

Since $\sigma$ is determined from the data itself, the choice of threshold used for the sigma stripping will impact the resulting noise measurements. The lower and more spread out the TOA S/Ns are, the more the choice of threshold will impact the results, as significantly more TOAs will be excised at lower thresholds. Therefore careful testing of different thresholds should be taken. For most of our pulsars we found no statistically significant difference when using thresholds between $2.5\sigma$ and $4\sigma$, but smaller thresholds removed over half of the TOAs, which would artificially bias our results to smaller values of pulse jitter, since we can directly measure \DISSerr.
 
\subsection{Pulse Jitter Sub-Band Correlations} \label{subbandcor}

Correlations between the frequency sub-banded residuals can also be indicative of pulse jitter \citep[][]{Shannon2012, Shannon2014,Lam2016a}. If the size of the sub-bands is $\gtrsim$\scintnu,  then the timing residuals will not be correlated across our band due to DISS. However, if the TOAs have a low S/N, then \temperr $\gg$ \jittererr, and any correlations will be undetectable, despite \temperr being uncorrelated across the band.

While the flux of the MSPs in our long observations are not unusual when compared to other observations of the same MSPs at the same frequencies \citep{Lam2019}, the steep spectral index that most pulsars exhibit \citep[][]{Jankowski2018} means that the S/N of these TOAs is much lower than similar observations at lower frequencies. To try to increase the S/N of the TOAs, we used sub-band widths of 400~MHz and integration times of 16~minutes. Since we obtained measurements of \scintnu for this epoch, we can be confident that these sub-band sizes are large enough to mitigate DISS correlations. While this yielded a two sub-band correlation with a small number of TOAs, it also increased the S/N of the TOAs used.

\subsection{Measuring DM Variations on Hour Timescales} \label{constrain_dm}

Delays due to dispersion by the ISM follow
\begin{equation} \label{dispersionEq}
    \Delta t \simeq 4.15 \times 10^{6} {\rm{ms}} \times {\rm{DM}} \left (\frac{1}{1500^{2}} - \frac{1}{\nu^{2}} \right),
\end{equation}
where $\Delta t$ is in ms, DM is in pc~cm$^{-3}$, $\nu$ is the frequency in MHz, and the delay here is referenced to a frequency of 1500~MHz. Previously \cite{Jones2017} studied DM variations in NANOGrav MSPs, including those in our sample, on timescales of days to years. They found that four MSPs, including PSR~J1614$-$2230, had DMs that varied on timescales less than 14 days. However, they were unable to probe variations on shorter timescales due to the cadence of their observations. However, individual NANOGrav observations could be used to study DM variations on timescales as short as $\sim1-30$ minute over 800~MHz bandwidth for the GBT or 600~MHz bandwidth for Arecibo.

The nature of our long observations allows us to look for DM variations on hour-long timescales along multiple lines of sight. Since NANOGrav observations are typically $\sim25$ minutes, we can obtain multiple DM measurements based on segments of equivalent length over the course of long observation, although over just a single frequency band. This allows us to look for DM variations in these MSPs on shorter timescales than have been studied before, but longer timescales than can be studied using the individual NANOGrav observations.

To do this, we split our observations into 32~minute integrations spanning 64~sub-bands (12.5~MHz per band), the same values used by NANOGrav \citep{Arzoumanian2018}, and fit for the DM in 32~minute sections with the \tempo\footnote{\sc{http://tempo.sourceforge.net}} pulsar timing package. The results are then visually inspected for apparent variations or evolution in time.

We can also estimate what the expected DM variations from the ISM for each MSP on the timescale of $\sim$hours will be using the same method as in \cite{Cordes2016}. In the strong scattering regime, the size of the scattering cone of the pulsar is much larger than the Fresnel scale,
\begin{equation}
    r_{F} = \sqrt{\frac{\lambda D}{2 \pi}}.
\end{equation}
Here $\lambda$ is the observing wavelength, here 1.5~GHz, and $D$ is the distance to the pulsar, given in Table \ref{ISSparams}, from the NANOGrav 11-yr parallax measurements. As the radio wave propagates through the ISM, the phase of the wave, $\phi$, is perturbed, causing a change in DM as
\begin{equation}
    \delta{\rm{DM}} = \frac{d\phi/d\nu}{\lambda {\rm{r_{e}}}},
\end{equation}
where r$_{e}$ is the classical electron radius. with an observing frequency of 1.5~GHz, we can convert $\lambda$r$_{e}$ in units of pc~cm$^{-3}$ to find $\delta{\rm{DM}} = 5.75 \times 10^{-8}$~pc~cm$^{-3}$ per radian of phase perturbation.

We can then use transverse velocity of the pulsar and the length of the observation to determine how many $r_{F}$ lengths it travels, and thus how many multiples of $\delta$DM we would expect the DM to vary by over the course of the observation. Here the transverse velocity is derived either from the proper motions measured in the NANOGrav 11-yr timing parameters or from the measured scintillation parameters as described in \cite{Cordes1998},
\begin{equation}
    V_{\rm{ISS}} = A_{\rm{ISS}} \frac{\sqrt{D \Delta\nu_{\rm{d}}}}{\nu \Delta t_{\rm{d}}},
\end{equation}
where $A_{\rm{ISS}} = 2.53 \times 10^4$~km~s$^{-1}$. Both transverse velocities are reported in Table \ref{ISSparams}. As stated in \cite{Cordes2016}, for most pulsars, the expected timescale for DM variations due purely to the ISM is $\sim$weeks, suggesting that we would not expected to see any DM variations over the course of our long observations. We report the expected $\delta$DM from the ISM over each of our individual MSP long observations in \S \ref{dmvars}.

\subsection{Timing Precision with Non-Contiguous TOAs} \label{ResidChunksMethods}

Our observations also offer an opportunity to test the precision of pulsar timing residuals when using a set of TOAs contiguous in time versus non-contiguous TOAs representing the equivalent length of time. This comparison is particularly useful for considering the impact of CHIME on precision pulsar timing, as it will be able to time many pulsars daily, but only for $\sim5$~minutes \citep{Ng2018}. Eq.~\ref{tempfittingerrs} breaks down at very low S/N \citep{Arzoumanian2015}, which may be the case for some pulsars in this mode of observing. For PTAs such as NANOGrav, increasing the cadence of pulsar observations will increase sensitivity to continuous wave sources and tracking rapid DM variations, but will still require high timing precision \citep{Lam2018d}. 

To test this precision, we folded each of our long observations into four minute integrations with 64 frequency channels (12.5~MHz per channel). We modeled these daily observations by requiring at least 30 minutes between TOAs. This has the advantage of mitigating correlated \DISSerr between the TOAs. We then compared the rms of these non-contiguous TOAs to the rms of an equivalent length of contiguous TOAs. The total contiguous length changes based on the overall length of the observation. As an example, for a 2.5 hour observation, if we require 30 minutes between each TOA, we will have five, non-contiguous, four minute integrated TOAs with 64 frequency channels, which will be compared to a contiguous observation of 20 minutes 
split into four-minute integrated TOAs and 64 frequency channels. 

To compare equivalent length observations, we found the rms of the residuals of the first contiguous set of TOAs. We then shifted the start of the contiguous observation in time by one TOA (here four minutes), and again calculated the rms of the residuals, and so on until we reached the end of each full long observation. 

For each set of noncontiguous TOAs we bootstrap sampled and calculated the rms of that bootstrapped set of residuals 10000 times. This bootstrapping allowed us to account for variations in TOA accuracy due to scintillation. We then shifted in time by one TOA (four minutes) and take each TOA separated by 30~minutes, bootstrap sampled and calculated the rms of the new set of residuals 10000 times, and so on until we exhausted all sets of TOAs that could be separated by 30~minutes. 

\section{Scintillation Parameters and Secondary Spectra} \label{scintparamsss}

Here we present and discuss the results of both our scintillation parameter analysis and our secondary spectra analysis. We were able to measure both scintillation parameters for most MSPs with both the 1-D and 2-D fitting methods, 
however, for PSRs~J0645$+$5158 and J1909$-$3744 we report only lower limits on \scintt. 
For all MSPs the 2-D method obtains more robust fits since there are many more points to fit despite the larger number of free parameters. The values we obtained for \scintnu compare well with the literature and predictions from NE2001, as do values of \scintt when using transverse velocity measurements from pulsar timing. Additionally we find evidence that for both PSRs~J0023$-$0923 and J1614$-$2230 the ISM differs from a purely uniform medium along the line of sight, and that for PSR~J1614$-$2230 a single scattering screen is insufficient to describe the ISM along the pulsars line of sight. We were unable to find any scintillation arcs in the secondary spectra by-eye.

\subsection{Scintillation Parameters Results and Discussion} \label{spresults}

\renewcommand{\arraystretch}{1.5}
\begin{deluxetable*}{l|cc|ccc|ccc}
\tablecaption{Fit Scintillation Parameters \label{fit_scint_params}}
\tablecolumns{9}
\tablehead{
\colhead{PSR} & \multicolumn{2}{c}{1-D Fitting} & \multicolumn{3}{c}{2-D Fitting} & \colhead{Levin et al. 2016} & \multicolumn{2}{c}{NE2001} \\ 
\colhead{} & \colhead{\scintt} & \colhead{\scintnu} & \colhead{\scintt} & \colhead{\scintnu} & \colhead{Drift Rate}  & \colhead{\scintnu} & \colhead{\scintt} & \colhead{\scintnu} \\
\colhead{} & \colhead{(minutes)} & \colhead{(MHz)} & \colhead{(minutes)} & \colhead{(MHz)} & \colhead{MHz/min} & \colhead{(MHz)} & \colhead{(minutes)} & \colhead{(MHz)}}
\startdata
J0023+0923& $51$ $\pm$ $10$ & $53$ $\pm$ $16$& $63$ $\pm$ $13$ &$ 50$ $\pm$ $15$ & $0.3$ $\pm$ $0.1$ & $21$ $\pm$ $6.6$ & $27^{+12}_{-7.0}$ & $42^{+34}_{-15}$\\ 
J0340+4130& $16$ $\pm$ $1$ & $3.7$ $\pm$ $0.2$& $12$ $\pm$ $1$ &$ 2.9$ $\pm$ $0.1$ & $0.03$ $\pm$ $0.01$ & $9.1$ $\pm$ $3.3$ & $22^{+16}_{-15}$ & $2.1^{+1.2}_{-0.8}$\\ 
J0613$-$0200& $11$ $\pm$ $1$ & $7.7$ $\pm$ $0.5$& $13$ $\pm$ $1$ &$ 3.6$ $\pm$ $0.2$ & $-0.01$ $\pm$ $0.01$ & $11$ $\pm$ $4$ & $14^{+8}_{-4}$ & $6.4^{+6.9}_{-2.6}$\\ 
J0645+5158 (1)& $>96$ & $67$ $\pm$ $26$& $>67$ &$ 80$ $\pm$ $31$ & $-0.6$ $\pm$ $0.2$  & -- & $29^{+12}_{-9}$ & $17^{+11}_{-6}$\\ 
J0645+5158 (2)& $>80$& $63$ $\pm$ $22$& $>140$  &$ 55$ $\pm$ $19$ & $-88$ $\pm$ $2$ & & & \\ 
J0645+5158 (3)& $>68$ & $53$ $\pm$ $18$& $>64$ &$ 593$ $\pm$ $339$ & $0.02$ $\pm$ $0.02$ & & & \\ 
J1614$-$2230& $12$ $\pm$ $1$ & $5.5$ $\pm$ $0.4$& $23$ $\pm$ $1$ &$ 4.9$ $\pm$ $0.5$ & $0.06$ $\pm$ $0.01$ & $9.0$ $\pm$ $2.6$ & $4.2^{+1.4}_{-0.8}$ & $3.6^{+2.4}_{-1.2}$\\  
J1832$-$0836& $4.8$ $\pm$ $0.1$ & $5.1$ $\pm$ $0.2$& $4.4$ $\pm$ $0.1$ &$ 4.2$ $\pm$ $0.2$ & $0.03$ $\pm$ $0.05$ & -- & $3.8^{+2.5}_{-2.0}$ & $5.9^{+5.6}_{-2.5}$\\ 
J1909$-$3744 (512)& $>82$ & $81$ $\pm$ $31$& $>62$ &$ 24$ $\pm$ $6$ & $25$ $\pm$ $3$ & $39$ $\pm$ $15$ & $13^{+5}_{-2}$ & $68^{+51}_{-24}$\\ 
J1909$-$3744 (256)& $>39$ & $9.9$ $\pm$ $1.9$& $>34$ &$ 26$ $\pm$ $7$ & $10.6$ $\pm$ $0.2$ & & & \\ 
J1909$-$3744 (192)& $>180$ & $88$ $\pm$ $45$& $>312$ &$ 300$ $\pm$ $188$ & $-0.4$ $\pm$ $0.2$ & & & 
\enddata
\tablecomments{Fitted scintillation parameters for our MSPs; all reported errors are $1\sigma$. The 1-D fitting was done in a similar way as described in \cite{Levin2016} where a single Gaussian is for each parameter after summing along the appropriate axis, however, we only summed the section of the 2-D ACF that was used in the 2-D fitting. The 2-D fitting was done by fitting a 2-D Gaussian to a subsection of the 2-D ACF. The values for \scintnu as found by \cite{Levin2016} and as calculated by \cite{NE2001} at center frequencies of 1500~MHz are reported for comparison. Errors for \scintnu from NE2001 come from the model outputs. The values for \scintt reported for the NE2001 model were computed using the \scintnu values from NE2001 but assuming $V_{\rm{ISS}} = V_{\perp}$ as calculated from the NANOGrav 11-yr timing parameters and reported in Table \ref{ISSparams}. Errors on these values of \scintt are propagated from the errors on $V_{\perp}$ and the pulsar distance from the NANOGrav parallax measurement.}
\end{deluxetable*}
\renewcommand{\arraystretch}{1}

\begin{deluxetable*}{lcccccc}
\centering
\tablecolumns{10}
\tablecaption{Scintillation-derived Parameters \label{ISSparams}}
\tablehead{
\colhead{PSR} & \colhead{$D$} & \colhead{$V_{\rm{ISS}}$}  &  \colhead{$V_{\perp}$} & \colhead{$\tau_{\rm d}$} &  \colhead{$n_{\rm{ISS}}$} & \colhead{\DISSerr}\\
\colhead{} & \colhead{(kpc)} & \colhead{(km s$^{-1}$)} & \colhead{(km s$^{-1}$)} & \colhead{(ns)} & \colhead{} & \colhead{(ns)} 
}
\startdata
J0023+0923 & $1.08\pm0.18$ & 33 $\pm$ 6 & $71 \pm 12$ & $3.2 \pm 1.0$ & $6 \pm 1$  &  $1.3 \pm 0.4$ \\ 
J0340+4130 & $1.4\pm0.9$ & 46 $\pm$ 14  & $22 \pm 14$ & $54.8 \pm 2.8$  & $240 \pm 10$ & $3.6 \pm 0.2$\\ 
J0613$-$0200 & $1.08\pm0.23$ & 44 $\pm$ 5 & $54 \pm 11$ & $43.6 \pm 2.3$ & $250 \pm 10$  &  $2.8 \pm 0.2$\\ 
J0645+5158 (1) & $1.22\pm0.28$ & 41 $\pm$ 9 & $44 \pm 10$ & $2.0 \pm 0.8$  & $4 \pm 1$  &  $1.1 \pm 0.4$\\ 
J0645+5158 (2) & & 16 $\pm$ 4 & & $2.9 \pm 1.0$ & $4 \pm 1$  &  $1.4 \pm 0.5$\\ 
J0645+5158 (3) & & 120 $\pm$ 40 & & $0.3 \pm 0.2$  &  $2 \pm 1$  &  $0.2 \pm 0.1$\\ 
J1614$-$2230 & $0.67\pm0.04$ & 22 $\pm$ 1 & $103 \pm 7$ & $32.7 \pm 3.2$ & $79 \pm 8$  &  $3.7 \pm 0.4$\\ 
J1832$-$0836 & $2.8\pm1.2$ & 220 $\pm$ 50 & $306 \pm 130$ & $37.9 \pm 1.4$  & $43 \pm 20$  &  $1.8 \pm 0.1$\\ 
J1909$-$3744 (512) & $1.09\pm0.04$ & 24 $\pm$ 3 & $191 \pm 6$ & $6.5 \pm 1.5$ & $10 \pm 2$  &  $2.1 \pm 0.5$\\ 
J1909$-$3744 (256) & & 44 $\pm$ 6 & & $6.1 \pm 1.7$ &  $7 \pm 2$  &  $2.4 \pm 0.7$ \\ 
J1909$-$3744 (192) & & 16 $\pm$ 5 & & $0.5 \pm 0.3$ & $1 \pm 1$  &  $0.5 \pm 0.3$\\
\hline
\hline
& $\Omega_{\mathrm{u},\Delta \nu_{\rm d}}$ &  $\Omega_{\mathrm{u},\Delta t_{\rm d}}$ & $d_s/D$ & $\Omega_{\rm scr}$ & $\theta_r$ & $t_{\rm geo,min}$ \\
& (mas$^2$) & (mas$^2$) &  & (mas$^2$) & (mas) & ($\upmu$s) \\
\hline
& \multicolumn{1}{|c}{} & \multicolumn{1}{c|}{} & & & & \\
J0023+0923 & \multicolumn{1}{|c}{$0.017\pm0.006$} & \multicolumn{1}{c|}{$0.004\pm0.002$} & $0.30\pm0.13$ & $0.014\pm 0.007$ & $0.22 \pm 0.15$ & $0.027\pm0.024$ \\
J0340+4130 & \multicolumn{1}{|c}{$0.23\pm0.14$} & \multicolumn{1}{c|}{$0.9\pm1.1$} & $0.89\pm0.13$ & $0.4\pm0.4$ & $0.18\pm0.19$ & $0.45\pm0.33$ \\
J0613$-$0200 & \multicolumn{1}{|c}{$0.24\pm0.05$} & \multicolumn{1}{c|}{$0.16\pm0.07$} & $0.57\pm0.12$ & $0.16\pm0.03$ & $-2.6\pm2.9$ & $12\pm24$ \\
J1614$-$2230 & \multicolumn{1}{|c}{$0.28\pm0.03$} & \multicolumn{1}{c|}{$0.013\pm0.002$} & $0.083\pm0.015$ & $0.61\pm0.16$ & $9.4\pm2.9$ & $6.4\pm2.8$ \\
J1832$-$0836  & \multicolumn{1}{|c}{$0.08\pm0.03$} & \multicolumn{1}{c|}{$0.04\pm0.03$} & $0.51\pm0.24$ & $0.05\pm0.02$ & $2.1\pm4.2$ & $20\pm60$ 
\enddata
\tablecomments{Values derived from the scintillation parameters resulting from the 2-D fits in Table \ref{fit_scint_params}. We have calculated $V_{\rm{ISS}}$, and compare that to $V_{\perp}$ from the NANOGrav 11-yr timing parameters. The distances to the pulsar $D$ from the NANOGrav parallax are also provided. Other parameters, $\tau_{\rm d}$, $n_{\rm{ISS}}$, and \DISSerr are directly related to determining the  white-noise contributions from the ISS. In the bottom section, we describe derived parameters from assuming the geometry of the medium along the line of sight to each pulsar is a uniform medium (scattering ``strength'' $\Omega_{\rm u}$ derived from \scintnu and \scintt, respectively), or a screen at a fractional distance from the Earth $d_s/D$ with strength $\Omega_{\rm scr}$. The scintle drift rates from the 2-D fits provide the refraction angle along the pulsar's direction of motion $\theta_r$ and the corresponding geometric time delay $t_{\rm geo,min}$ given the screen distance.}
\end{deluxetable*}

Using both methods described in \S \ref{scintparamsmethods} we have determined scintillation timescales and bandwidths for each of the MSPs observed. The different parameters for each method for each MSP can be found in Table \ref{fit_scint_params}.

In general the 1-D parameter fitting agrees with the 2-D parameter fitting within $3\sigma$. However, when summing over one axis, excess noise may be added to the 1-D ACF, making it difficult to fit a single Gaussian, whereas in the 2-D fit, there are many more samples being fit over, minimizing the excess noise. The 2-D Gaussian fit scintillation parameters also match previously obtained results in the literature within $3\sigma$ in all cases. We can also analyze the scintle drift rates only by using the 2-D fits. Due to these factors, we have used the 2-D fit scintillation parameters for the remainder of this work.

For all of the binary MSPs analyzed in this work, the observation lengths are generally much shorter than the binary orbital periods, with the exception of PSR~J0023$+$0923. We did not see any orbital-phase dependent changes in the scintillation properties as seen in some binary pulsars \citep{Rickett2014, Reardon2019}.

\subsubsection{Scintillation Bandwidths}

When compared to \cite{Levin2016}, the values obtained for \scintnu in this work match within 2$\sigma$. The discrepancies in these values are likely due to the fact that while the values of \scintnu obtained in \cite{Levin2016} were averaged over many epochs, the values reported in this work relate to a single epoch. As the ISM is a dynamic environment it is expected that the measured values of \scintnu will change in time, so some differences are expected. Additionally, our measurements of \scintnu better match those from \cite{Levin2016} for MSPs with fewer scintles. This may be because with a longer observation more scintles will be observed than in a shorter observation, but for MSPs with large \scintt and \scintnu, the numbers may be comparable. However, as we have a larger $n_{\rm{ISS}}$ in almost all cases we expect our parameters to be more accurate than those found in \cite{Levin2016}.

For PSR~J1909$-$3744, the value of \scintnu from the observation with only 192 frequency channels (300~MHz bandwidth) deviates greatly from the other values, spanning the full bandwidth of the observation. This discrepancy is likely due to both the smaller bandwidth and the lack of scintles observed in frequency for this section of the observation. 

We report values of \scintnu for PSRs~J0645$+$5158 and J1832$-$0836 for the first time in this work. For the two sections of the first observation of PSR~J0645$+$5158, (1) and (2), the values of \scintnu agree with each other within $1\sigma$. The third observation, (3), taken 10~days later, has a \scintnu almost spanning the full bandwidth, and different from the other two measurements by a factor of $\sim12$. However, from the dynamic spectrum shown in Figure \ref{DynSpec_Panels}, we can see that for this observation we appear to only resolve a single scintle spanning the full bandwidth and length of the observation. The differences in \scintnu are therefore not surprising, and the actual value is likely closer to that of the first two segments. For J1832$-$0836, \scintnu is quite small, only about three times our frequency resolution. 

When compared to the predicted values of \scintnu from NE2001, all of the values obtained from our long observations are within 2$\sigma$. While there is some variation, this shows the accuracy of the NE2001 model in predicting \scintnu along different lines of sight.

\subsubsection{Scintillation Timescales}

We also report values or lower limits of \scintt for all seven MSPs for the first time. We note that our observation of PSR~J0023$+$0923 contains only a small number of scintles (see Table \ref{ISSparams}). Additionally, due to the observing issues and length of the observations for both PSRs~J0645$+$5158 and J1909$-$3744, there were no fully time resolved scintles, and the values reported here are at best lower limits on \scintt. For these two MSPs the \scintt lower limit reported from 1-D fitting is the length of the observation.

For the NE2001 \scintt estimates, we have calculated \scintt using the same method as \cite{NE2001}, but have have used the proper motion velocity $V_\perp$ as determined from the NANOGrav 11-yr timing parameters as a proxy for $V_{\rm{ISS}}$. These values are reported in Table \ref{fit_scint_params} and match the derived values of \scintt within $2\sigma$ for all MSPs except PSRs~J1614$-$2230 and J1909$-$3744. While it has been shown that $V_{\rm{ISS}}$ and $V_\perp$ closely follow each other, differences in the two velocities can occur if the scattering does not occur uniformly along the line of sight \citep{Lyne1982, Cordes1986}. Further analysis of these differences is beyond the scope of this work and left for future analyses.

\subsubsection{Uniform Media vs Thin Screens} \label{screen}

Following Appendix C of \citet{Cordes1998}, we can use the scintillation bandwidth and timescale to constrain the properties of the medium along the line of sight. We tested two geometries: a uniform medium and a thin scattering screen. For the latter, we can determine the distance and ``strength'' of the screen uniquely. The two parameters are related to these via
\begin{eqnarray}
\tau_{\rm{d}} & = & \frac{\eta_0 \Delta s}{2c} d_s \left(1 - \frac{d_s}{D}\right),\\
\Delta t_{\rm{d}} & = & \frac{\lambda}{\pi V_\perp} \left(\frac{1}{2 \eta_0 \Delta s}\right)^{1/2} \frac{D}{d_s},
\end{eqnarray}
where $\lambda = c/\nu$ is the electromagnetic wavelength and again $d_s$ is the Earth-screen distance. Recall that we can relate \scintnu to \scinttau via Eq.~\ref{scateq}. We define the strength of the screen $\Omega_{\rm scr} \equiv \eta_0 \Delta s$ which is the product of the mean-square scattering angle per unit length along the line of sight times the thickness of the screen. The electron-density wavenumber spectrum has an amplitude proportional to $\eta_0$ \citep{Cordes1998}, and therefore when multiplied by the screen thickness $\Delta s$ gives the integrated scattering strength. The units of $\eta_0$ are often written in mas$^2$ kpc$^{-1}$ and $\Omega$ has units of mas$^2$.

The scintillation parameters in the uniform medium case take a simpler form,
\begin{eqnarray}
\tau_{\rm{d}} & = & \frac{\eta_0 D^2}{2c},\\
\Delta t_{\rm{d}} & = & \frac{\lambda}{\pi V_\perp} \left(\frac{3}{2 \eta_0 D}\right)^{1/2}.
\end{eqnarray}
For comparison with the thin-screen case, we can define $\Omega_{\rm u} \equiv \eta_0 D$ to give the comparable scattering strength over the entire line of sight. If the $\Omega_{\rm u}$ obtained from both scintillation parameter measurements is consistent, then scattering is consistent with coming from throughout the line of sight.

Using the five MSPs in which we constrained both \scintt and \scintnu, we calculated the separate $\Omega$ values obtained for a uniform medium as well as the $d_s/D$ and $\Omega_{\rm scr}$ values for the thin-screen geometry. These solutions are provided in  Table~\ref{ISSparams}. We see that for PSRs~J0023+0923 and J1614$-$2230, the $\Omega$ values derived from the scintillation bandwidths and timescales are not consistent with each other, suggesting the line of sight differs from a purely uniform medium. These two MSPs have the tightest constraints on the $\Omega$ values, so it is possible that the lines of sight for all five of these MSPs differ slightly from a uniform medium, but we are unable to constrain them well enough to verify this.

\subsubsection{Scintle Drift Rates} \label{rots}

As discussed, we obtained scintle drift rates for the five MSPs in which we performed the 2-D ACF fitting. We note that for PSR~J1614$-$2230, we had to constrain the time-frequency space for the 2-D ACF fit. As shown in Figure~\ref{DynSpec_Panels}, it appears that the drift rates are negative, or that the scintles are moving from higher to lower frequencies in time, in the first half of the observation at higher frequencies (with timescale $\gtrsim$1 hour). This resulted in a smaller scale peak in the ACF with a positive drift rate, where the scintles appear to move from lower to higher frequencies with time, on top of a much larger feature with a negative drift rate. Upon closer inspection, we found that what appears to be two scintles in the top left of the dynamic spectrum (the beginning of the observation at higher frequencies) are actually several bright scintles with similar drift rates to the other scintles throughout the observation. This shows that a large number of scintles are required to measure scintillation parameters without a systematic bias; if we had a smaller bandwidth and/or a shorter observation time, we would have measured the scintillation parameters incorrectly. Since we believe that the shorter timescale is more representative of the characteristic scintillation timescale (\scintt $\sim$~23~min), we constrained the 2-D Gaussian to fit only over the central portion of the ACF. For the other pulsars, we visually inspected the fits to ensure we were unbiased in our measurements and did not note another instance of this apparent drifting.

From the drift rates in the scintles, we can derive the refraction angle $\theta_R$ along the pulsar's direction of motion using Eq.~\ref{eq:driftrate}. Since the refracted emission takes longer to travel to the pulsar, it is associated with a geometric time delay equal to \citep{Cordes2010,Lam2016b}
\begin{equation}
    t_{\rm geo,min} = \frac{1}{2c}\left(\frac{D d_s}{D-d_s}\right) \theta_r^2.
\end{equation}
This is the minimum delay since we do not know the refraction angle in the direction perpendicular to the pulsar motion. Both $\theta_r$ and $t_{\rm geo,min}$ are given in Table~\ref{ISSparams}. Measurement of these geometric delays is critical; since the refraction delay is $\propto \nu^{-4}$, removal of the dispersive $\nu^{-2}$ delay will bias the ``infinite-frequency'' arrival times used in precision timing experiments \citep{Lam2016b,Lam2018b}. We found $t_{\rm geo,min}$ was roughly consistent with zero for all pulsars after propagation of all uncertainties except for PSR~J1614$-$2230. Assuming the delay is purely refractive and $\propto \nu^{-4}$, the value of $t_{\rm geo,min} = 6.4\pm2.8~\upmu$s implies a $\sim70~\upmu$s delay at the 820~MHz band also used by NANOGrav, and therefore a DM perturbation of amplitude $-$0.015~pc~cm$^{-3}$ \citep{Lam2018a}, which is unseen in NANOGrav data \citep{Arzoumanian2018}. We therefore believe that $t_{\rm geo,min}$ is biased and neither a uniform medium nor single scattering screen are adequate to describe this line of sight; a more in-depth analysis on this pulsar's varying scintillation parameters using NANOGrav data will be performed in future work (M.~T.~Lam et al., in prep.).

\subsubsection{Estimating \DISSerr}

From the fit values of \scintt and \scintnu we calculated values of $\tau_{\rm{d}}$, $\sigma_{\rm{DISS}}$, and $n_{\rm{ISS}}$ from Eqs. \ref{scateq}, \ref{DISSeq}, and \ref{nscintles} respectively, which are reported in Table \ref{ISSparams}. For all MSPs where we are able to resolve \scintt, we report the most accurate values for $n_{\rm{ISS}}$ and $\sigma_{\rm{DISS}}$ for this epoch. For MSPs where \scintt is a lower limit, we took $(1 + \eta_{t} (T / \Delta t_{\rm{d}})) \approx 1$, as is typically done \cite[][]{Levin2016}. We found that for all MSPs, $\sigma_{\rm{DISS}}$ is on the order of nanoseconds, and smaller than or equal the values of \DISSerr found by \cite{Lam2016a} and are thus a very small contribution to the white noise present in the timing residuals. The large number of scintles present in our long observations along with our new scintillation parameters likely account for the smaller \DISSerr value.

\subsection{Secondary Spectra Results and Discussion} \label{secspecresults}

No arcs were visually apparent in any of our observations, and therefore we did not further analyze the secondary spectra. While some pulsars, such as PSR~J1614$-$2230, appear to be slightly brighter on one side of the zero conjugate frequency, there is no clear evidence of scintillation arcs. This could be due to a lack of frequency resolution in the dynamic spectra. However, with no obvious detection of any arcs, further analysis of the secondary spectra is beyond the scope of this work.

\section{Pulse Jitter Results and Discussion} \label{allthejitter}

Here we will present and discuss the results of our measurements of pulse jitter. We have attempted to measure the pulse jitter contribution to the white noise in our timing residuals using a direct method by fitting values of \residerr as a function of integration time as well as using a maximum likelihood analysis. While values of \jittererr can be determined with both methods, in general, our TOAs do not have a high enough S/N to separate out the \jittererr contribution to the white noise. As a final test, we discuss the results of our sub-band correlation analysis, which is similarly hindered, despite a large sub-band width and integration time.

\begin{figure*}
\includegraphics[width = \textwidth ]{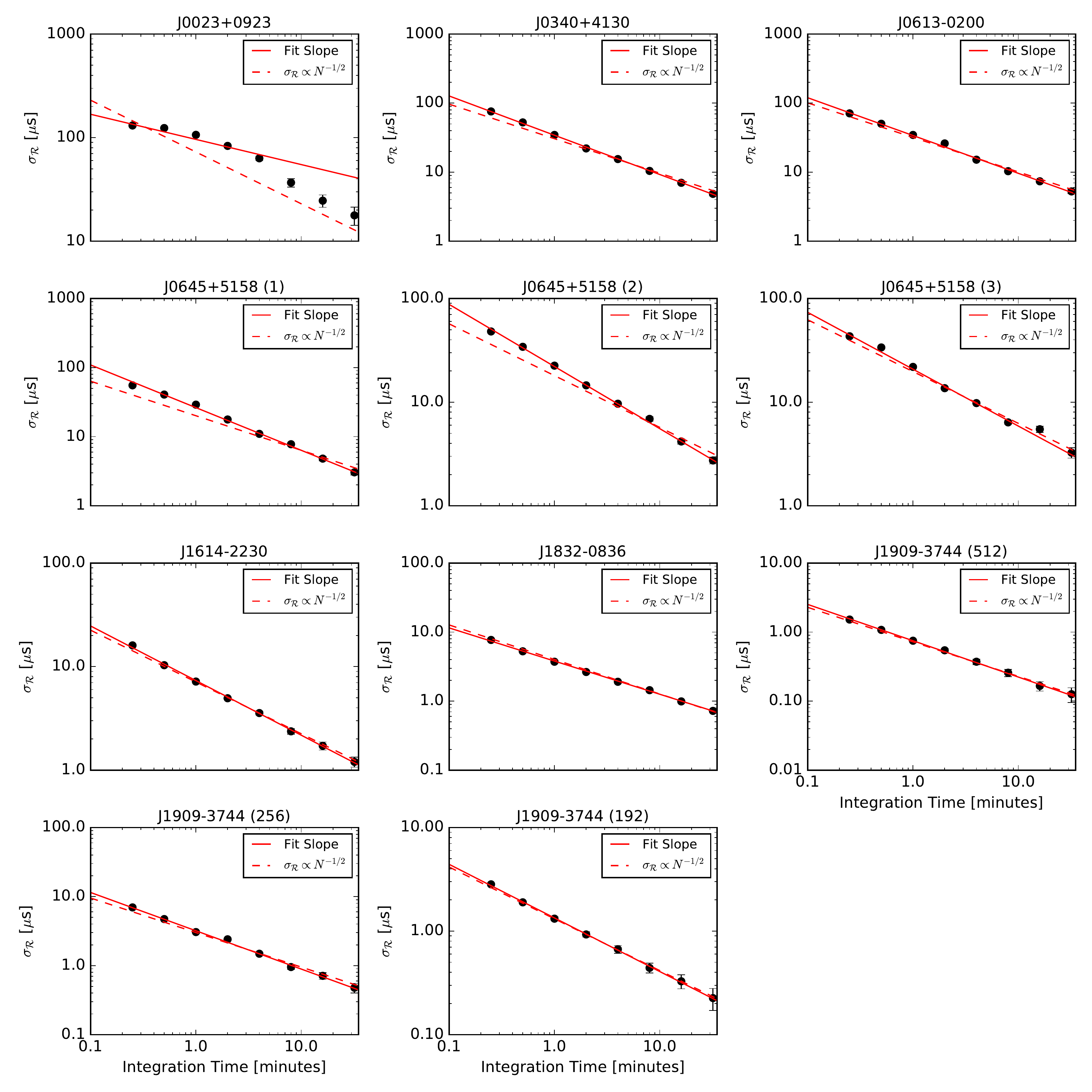}
\caption{\residerr as a function of integration time calculated from the residuals generated with \pypulse for all MSPs analyzed in this work in log space. All points are potted with error bars, but not all are visible. The dashed red line shows \residerr $\propto N^{-1/2}$, where $N$ is the number of pulses in each folded integration and fit for just \residerr for single pulses. This is expected if we assume that all pulses from the pulsar are statistically independent. The solid line fits for both \residerr for single pulses and a dependence on $N^{\alpha}$, where $\alpha$ is the slope of the line in log space. The fit values for both of these lines are reported in Table \ref{resid_sublen_vals}. \label{JitterAnalysis1}}
\end{figure*}

\subsection{Pulse Jitter Meausured from $\sigma_{\cal{R}}$} \label{jitterresults}

\begin{deluxetable*}{l | c | c c | c c}
\tablecaption{\residerr and Integration Time Fitting  \label{resid_sublen_vals}}
\tablecolumns{6}
\tablehead{
\colhead{PSR} & \colhead{Fixed \jittererr} & \colhead{Fit $\alpha$} & \colhead{Fit \jittererr} & \colhead{F statistic} & \colhead{F Significance} \\
\colhead{} & \colhead{(ms)} & \colhead{} & \colhead{(ms)} & \colhead{} & \colhead{}}
\startdata
J0023+0923 & $10.1 \pm 0.2$ & $-0.25 \pm 0.05$ & $13.4 \pm 0.2$ & 37.3 & 0.0009 \\ 
J0340+4130 & $4.1 \pm 0.1$ & $-0.57 \pm 0.01$ & $4.6 \pm 0.1$ & 11.0 & 0.0162 \\ 
J0613$-$0200 & $4.5 \pm 0.1$ & $-0.55 \pm 0.01$ & $4.8 \pm 0.1$ & 5.8 & 0.0526 \\ 
J0645+5158 (1) & $1.7 \pm 0.1$ & $-0.62 \pm 0.01$ & $2.2 \pm 0.1$ & 23.1 & 0.0030 \\ 
J0645+5158 (2) & $1.5 \pm 0.1$ & $-0.60 \pm 0.01$ & $1.8 \pm 0.1$ & 14.5 & 0.0089 \\ 
J0645+5158 (3) & $1.6 \pm 0.1$ & $-0.55 \pm 0.02$ & $1.7 \pm 0.1$ & 13.3 & 0.0108 \\ 
J1614$-$2230 & $1.0 \pm 0.1$ & $-0.53 \pm 0.01$ & $1.0 \pm 0.1$ & 7.7 & 0.0321 \\ 
J1832$-$0836 & $0.6 \pm 0.2$ & $-0.48 \pm 0.01$ & $0.6 \pm 0.2$ & 35.1 & 0.0010 \\ 
J1909$-$3744 (512) & $0.1 \pm 0.2$ & $-0.53 \pm 0.01$ & $0.1 \pm 0.2$ & 12.7 & 0.0119 \\ 
J1909$-$3744 (256) & $0.4 \pm 0.2$ & $-0.55 \pm 0.01$ & $0.5 \pm 0.2$ & 4.6 & 0.0763 \\ 
J1909$-$3744 (192) & $0.2 \pm 0.1$ & $-0.52 \pm 0.01$ & $0.2 \pm 0.1$ & 11.9 & 0.0136 
\enddata
\tablecomments{Estimated single pulse jitter values based on fitting for \residerr as a function of the integration time. Fixed values assume that \residerr $\propto N^{-1/2}$, the number of pulses in the integration. Fit values are fitting for this dependency as \residerr $\propto N^{\alpha}$. The F statistic and significance compare how significant fitting for the $\alpha$ value is to the fit. We use a significance of 0.0027 (3$\sigma$) to determine if fitting for $\alpha$ is significant. We find that it is significant for only two MSPs, PSRs~J0023$+$0923 and J1832$-$1836.}
\end{deluxetable*}

Our first attempt to measure the pulse jitter is from  direct calculation using Eq. \ref{whitenoiseerrs} with \residerr calculated as a function of integration time and using measured values of both \temperr and \DISSerr. Figure \ref{JitterAnalysis1} shows the two different fits of \residerr, with the $1/N^{\alpha}$ represented by the solid red line, and the $1/\sqrt{N}$ by the dashed red line. We performed an F-test to calculate the significance of fitting a varying slope compared to a constant $\sqrt{N}$ to these \residerr as a function of integration time with a significance threshold of 0.0027 (3$\sigma$); the F-statistic and its significance are reported in Table \ref{resid_sublen_vals}.

We find that fitting a slope instead of assuming a fixed $\sqrt{N}$ is significant for only two MSPs, PSRs~J0023$+$0923 and J1832$-$0836. However, for PSR~J1832$-$0836, $\alpha = -0.48\pm 0.01$, which is within $2\sigma$ of the expected value of $-0.5$. For PSR~J0023$+$0923, the shallower slope that is fit for shorter integration times is indicative that the individual pulses may not be statistically independent \citep{Helfand1975, Rankin1995}. Despite this apparently significant fit, we found that \residerr for a single pulse is consistent within 1$\sigma$ whether we fit for the slope or not. We will therefore discuss just the values obtained when assuming that \residerr $\propto 1/\sqrt{N}$ for the remainder of this work.

We report the values of \residerr, \temperr, and \jittererr for both two minute integrations with 64 frequency channels (12.5~MHz per channel) as well as the values extrapolated back for single pulses. For \temperr we took the value of $S$ in Eq. \ref{tempfittingerrs} to be the median value of $S$, calculated as the peak to off-pulse rms ratio, for all the TOAs used in each set of integrations. For some MSPs this results in \temperr $>$ \residerr. If no value is reported then we were unable to determine a value of \jittererr\ with this method. However, this result in general shows that we cannot assume any TOAs from a given observation will be in the high S/N regime, meaning \temperr $\gg$ \jittererr.

\begin{deluxetable*}{l | c c c | c c c | c c c}
\tablecaption{\jittererr Estimates from Fitting \label{Fit_jitter}}
\tablecolumns{10}
\tablehead{
\colhead{PSR} & \colhead{$W_{\rm{eff}}$} & \multicolumn{2}{c}{Max Likelihood} & \multicolumn{3}{c}{2 Minute Integrations} & \multicolumn{3}{c}{Single Pulse} \\
\colhead{} & \colhead{}& \colhead{$\sigma_{C}$} & \colhead{\jittererr}  & \colhead{$\sigma_{\cal{R}}$} & \colhead{$\sigma_{\rm{S/N}}$} & \colhead{$\sigma_{\rm{J}}$} & \colhead{$\sigma_{\cal{R}}$} & \colhead{$\sigma_{\rm{S/N}}$} & \colhead{$\sigma_{\rm{J}}$} \\
\colhead{} & \colhead{($\upmu$s)} & \colhead{($\upmu$s)} & \colhead{($\upmu$s)} & \colhead{($\upmu$s)} & \colhead{($\upmu$s)} & \colhead{($\upmu$s)} & \colhead{(ms)} & \colhead{(ms)} & \colhead{(ms)}
}
\startdata
J0023+0923 & 430 & $ 9.6^{+3.0}_{-3.2}$ & $ 9.6 $ &$ 83$ $\pm$ $2$ & $43$ $\pm$ $46$ & $71$ & $10.1$ $\pm$ $0.2$ & $8.6$ & $5.4$ \\
J0340+4130 & 517 & $ 4.8^{+0.9}_{-0.9}$ & $ 4.8 $ &$ 22.1$ $\pm$ $0.4$ & $23$ $\pm$ $13$ & -- & $4.1$ $\pm$ $0.1$ & $4.4$ & -- \\
J0613$-$0200 & 332 & $ 5.4^{+0.8}_{-0.8}$ & $ 5.4 $ &$ 26.1$ $\pm$ $0.5$ & $24$ $\pm$ $18$ & $11$ & $4.5$ $\pm$ $0.1$ & $4.7$ & -- \\ 
J0645+5158 (1) & 633 & $ 7.4^{+0.8}_{-0.8}$ & $ 7.4 $ &$ 17.7$ $\pm$ $0.4$ & $16$ $\pm$ $8$ & $8$ & $1.7$ $\pm$ $0.1$ & $1.8$ & -- \\ 
J0645+5158 (2) & 633 & $ 5.8^{+0.6}_{-0.6}$ & $ 5.8 $ &$ 14.5$ $\pm$ $0.3$ & $13$ $\pm$ $7$ & $6$ & $1.5$ $\pm$ $0.1$ & $1.5$ & -- \\ 
J0645+5158 (3) & 633 & $ 4.5^{+0.5}_{-0.5}$ & $ 4.5 $ &$ 13.6$ $\pm$ $0.3$ & $12$ $\pm$ $5$ & $6.6$ & $1.6$ $\pm$ $0.1$ & $1.4$ & $0.8$ \\ 
J1614$-$2230 & 403 & $ 4.9^{+0.2}_{-0.2}$ & $ 4.9 $ &$ 5.0$ $\pm$ $0.2$ & $6$ $\pm$ $5$ & -- & $1.0$ $\pm$ $0.1$ & $1.2$ & -- \\ 
J1832$-$0836 & 188 & $ 0.7^{+0.1}_{-0.1}$ & $ 0.7 $ &$ 2.6$ $\pm$ $0.1$ & $3$ $\pm$ $2$ & -- & $0.6$ $\pm$ $0.2$ & $0.7$ & $0.2$ \\ 
J1909$-$3744 (512) & 266 & $ 0.1^{+0.1}_{-0.1}$ & $ 0.1 $ &$ 0.6$ $\pm$ $0.1$ & $0.9$ $\pm$ $0.8$ & -- & $0.1$ $\pm$ $0.2$ & $0.2$ & $0.2$ \\ 
J1909$-$3744 (256) & 266 & $ 1.6^{+0.2}_{-0.2}$ & $ 1.6 $ &$ 2.4$ $\pm$ $0.1$ & $2.7$ $\pm$ $1.7$ & -- & $0.4$ $\pm$ $0.2$ & $0.5$ & $0.2$\\ 
J1909$-$3744 (192) & 266 & $ 0.1^{+0.1}_{-0.1}$ & $ 0.1 $ &$ 0.9$ $\pm$ $0.1$ & $1.5$ $\pm$ $1.4$ & -- & $0.2$ $\pm$ $0.1$ & $0.3$ & $0.2$ 
\enddata
\tablecomments{Estimates of the pulse jitter from the fit of the rms of the timing residuals as shown in Figure \ref{JitterAnalysis1} and compared with the results from the maximum likelihood analysis. The values of $W_{\rm{eff}}$ and \DISSerr are the same as reported in Table \ref{ISSparams}. The values used for \temperr and \residerr are reported for both the two minute integration residuals and the values extrapolated for single pulses. If no value is reported for \jittererr, this method is unable to estimate \jittererr. Maximum likelihood values are taken for 50~MHz wide frequency channels and two minute integrations.}
\end{deluxetable*}

\subsection{Maximum Likelihood Jitter Results} \label{funneljitter}

Our second attempt at estimating the pulse jitter from the maximum likelihood analysis also found results similar to those presented above. While a statistically significant value for the pulse jitter was determined using this analysis, \temperr was found to be larger than \jittererr in all cases reinforcing that our high S/N regime assumption does not hold for these observations. Despite the limiting S/N of our TOAs, the results of the maximum likelihood pulse jitter analysis are reported in Table~\ref{Fit_jitter}. 

The maximum likelihood analysis found that in all cases the values of \sigmac are significant and are not 95\% upper limits. Using our measurements of \DISSerr we could separate out the \jittererr from \sigmac. However, since \sigmac $\gg$ \DISSerr as estimated from the dynamic spectra in all cases, the majority of the contribution to \sigmac appears to be from the pulse jitter. We note that all values of \jittererr found here are much larger than those found by \cite{Lam2016a} and \cite{Lam2019} and are not representative of the best constraints that may be placed on the pulse jitter of these seven MSPs. Even so, using a maximum likelihood analysis better constrains \jittererr than directly solving Eq. \ref{whitenoiseerrs}.
 
 While our values of pulse jitter from the maximum likelihood analysis are larger than expected, \cite{Lam2016a} used the full NANOGrav 9-yr data set in their analysis, and \cite{Lam2019} used the full NANOGrav 12.5-yr data set, which allows for many more TOAs than our single long observations. The longer 9-yr/12.5-yr data set also means there is a greater chance of observing the pulsar during a particularly bright DISS time, or during a period of strong refractive scintillation (RISS), which can increase the observed flux density of the pulsar by factors of $\sim$2 \citep{Stinebring2000}. 
 
 The typical timescale of RISS is typically days to weeks, increasing with pulsar distance \citep{Sieber1982, Rickett1984, Hancock2019}. With just one epoch of observation, even spanning many hours, we are unlikely to have observed during a period of strong RISS for any pulsar when compared to many observations spanning multiple epochs. This, in addition to the low likelihood of observing during a period of bright DISS, are likely the primary explanations for the lower S/N of our observations and our ability to put limits on the pulse jitter.

\subsection{Sub-band Correlation Results} \label{subbandjitter}

We expect that at high S/N we will be able to see pulse jitter correlated across frequency channels as in \cite{Shannon2012, Shannon2014}. However, even with 16~min integrations and two frequency channels of 400~MHz each, the correlations between the TOAs in the two sub-bands are minimal. The correlation coefficients are small which is indicative of not being able to detect the pulse jitter in our observations.

This lack of TOA correlations between sub-bands is not surprising given the low pulse S/Ns discussed in \S \ref{funneljitter}. Since we must be in the high S/N limit to see indications of pulse jitter in the sub-band correlations, we do not meet our initial assumptions necessary for this analysis. 

\begin{figure*}[p]
    \centering
    \includegraphics[width=\textwidth]{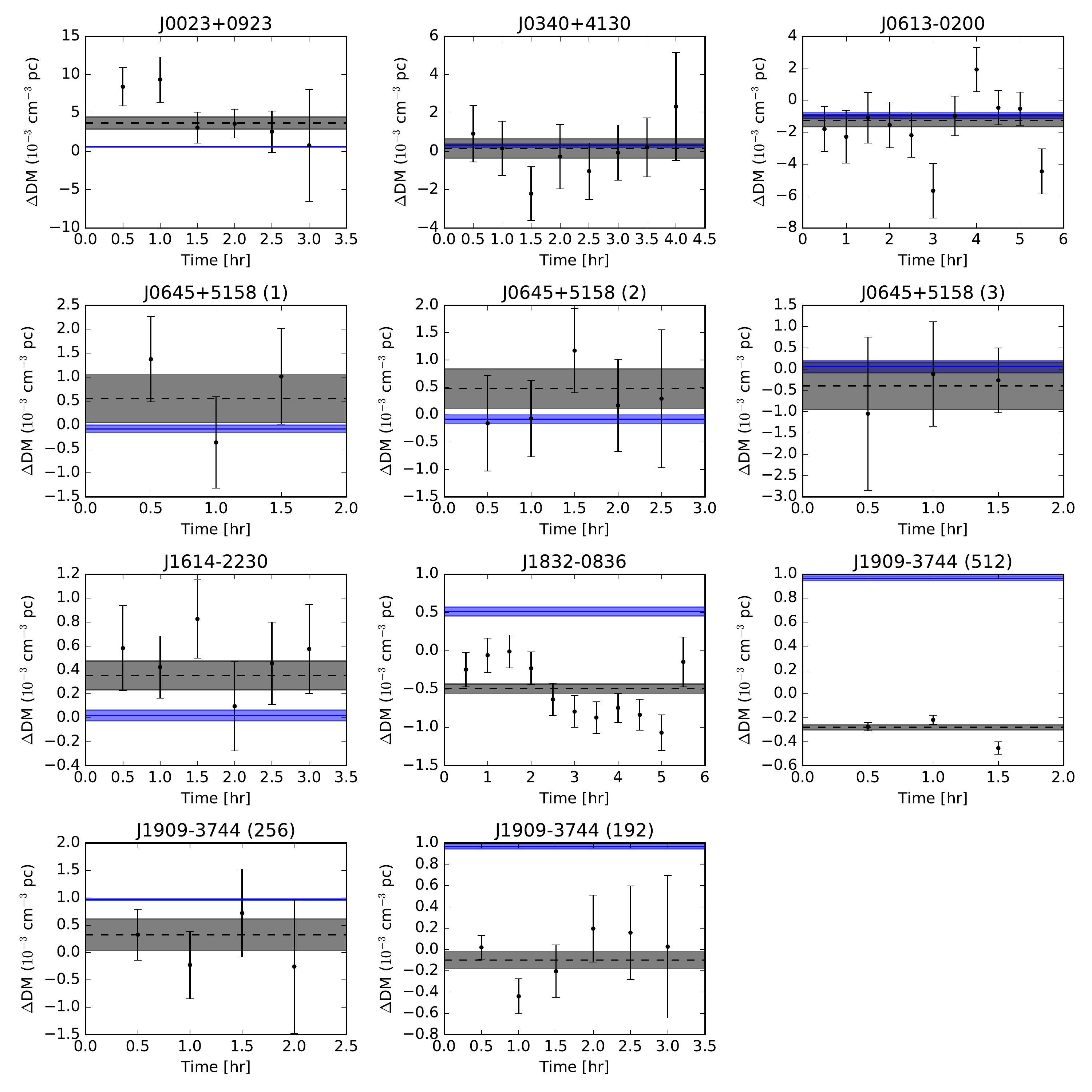}
    \caption{DM variations referenced to the nominal DM value from the NANOGrav 11-yr timing parameters and reported in Table \ref{mspinfo} over the course of each long observation analyzed in this work. The blue line is the DM value from the NANOGrav 11-yr data set from the same or closest epoch to our observations and the blue filled region shows the 1$\sigma$ errors. The black dashed line is the DM value fit from \tempo over the full long observation where the black filled region shows the $1\sigma$ errors also from \tempo. The black points are the fit DM value of using 32~minute integrations of the long observation with $1\sigma$ errorbars from \tempo. Three panels are shown for PSR~J1909$-$3744 because the observation was split into three sections of different bandwidths due to data acquisition instrumental difficulties when recording the data. Differences between the fit values in this work and those from NANOGrav are likely due to the larger frequency band used to fit the DM in the NANOGrav data. In most cases there is little variation over the course of the observation and all fit $\Delta$DM values are consistent within 2$\sigma$.}
    \label{All_DM_Vars}
\end{figure*}

\section{DM Variations on Short Timescales} \label{dmvars}

We found that for almost all MSPs analyzed, the DM is consistent within $1\sigma$ throughout the observation, and all points are consistent within $2\sigma$, as expected from \cite{Jones2017}. DM variations referenced to the nominal DM value from the NANOGra 11-yr timing parameters and reported in Table \ref{mspinfo} are shown in each panel in Figure \ref{All_DM_Vars}. For all MSPs our fit DM value differs from the value in the NANOGrav 11-yr data set at the same or closest epoch on the order of $\sim10^{-3}$~pc~cm$^{-3}$ or less.

Here we have taken our $1\sigma$ uncertainties directly from the \tempo fitting. We note that the S/N of the pulses will vary across the frequency band and with time due to scintillation. However, for PSRs~J0340$+$4130, J0613$-$0200, J1614$-$2230, and J1832$-$0836, both \scintt and \scintnu are smaller than the integration time of 32~minutes and frequency channel width of 12.5~MHz, so we expect the S/N to be roughly the same for each TOA. For the other MSPs, we expect that any variations in S/N across the band or with time should be accounted for with larger or smaller uncertainties on the fit DM.

Using the estimation technique described in \S \ref{constrain_dm}, we have also determined what the expected variation in DM should be for each MSP. To do this we have used only the longest observation length for each MSP if they have multiple observations, and have used $V_{\rm{ISS}}$ for all MSPs where we were able to measure \scintt, and $V_{\perp}$ otherwise. With distances between 0.67 and 2.4~kpc, observation lengths of 2.46 to 5.76 hours, and transverse velocities of 22 to 220~km~s$^{-1}$, we find that the DM variations expected from the ISM for the MSPs analyzed here range between $2-15 \times 10^{-8}$~pc~cm$^{-3}$ assuming a frequency of 1500~MHz. This is significantly smaller than any variations we can measure, and reinforces our expectation that the DM will not vary on $\sim$hour-long timescales.

The difference between our fit DM and the NANOGrav 11-yr DM values is likely due to the fact that while we are fitting for a single DM value over the full bandwidth at one epoch, the NANOGrav DM values are fit using six day bins which often include additional observations taken at other frequency bands \citep{Arzoumanian2015}. Our smaller frequency range likely biases our DM fit which would account for the difference between the two DM values. This shows the importance of fitting the DM over as large a frequency range as possible.

While most fit DM value are within $1\sigma$ of the expected DM value from our full observation DM fit, some fall $2\sigma$ away. One possible explanation for the larger DM differences on these short timescales could be changes in the ionosphere. However, \cite{Lam2016b} has shown these variations to be on scales much smaller than our fit DM differences. It is also possible that variations in the pulse profile over the course of our observation could cause these variations. However, \cite{Brook2018} have shown that for the MSPs presented in this work the profile variations are very small and therefore unlikely sources of these variations.

Using Eq. \ref{dispersionEq} we can find what the peak pulse profile shift in time would need to be to account for the DM difference in the 32~minutes integrations and the full observation DM fit. For the largest DM difference we find, $0.0056$~pc~cm$^{-3}$ for PSR~J0023$+$0923, we find a shift of 10~$\upmu$s from the expected TOA would be required to explain the difference in DM. For this MSP, our observation covers about 85\% of its binary orbit. However, using the NANOGrav 11-yr data set we find no correlation between the DM variations and orbital phase for PSR~J0023$+$0923, so we do not believe that the variations we find here are due to the orbital phase of the MSP.

For most other $2\sigma$ DM differences we find, peak pulse profile shifts on the order of 1~$\upmu$s are required, which is of the order of the timing precision of most NANOGrav MSPs \citep{Arzoumanian2018} and therefore variations at this level are expected. Intrinsic variations in the pulse profile with frequency or variations due to scattering may also account for these shifts.

As no MSPs in this work show DM variations larger than $2\sigma$ from the expected DM value over each observation, which can be accounted for as discussed above, we conclude that the DM of these MSPs does not vary on hour-long timescales.

\section{Timing Precision of Non-contiguous Timing Residuals Results and Discussion} \label{ResidChunkResults}

\begin{figure*}[p]
    \centering
    \includegraphics[width=\textwidth]{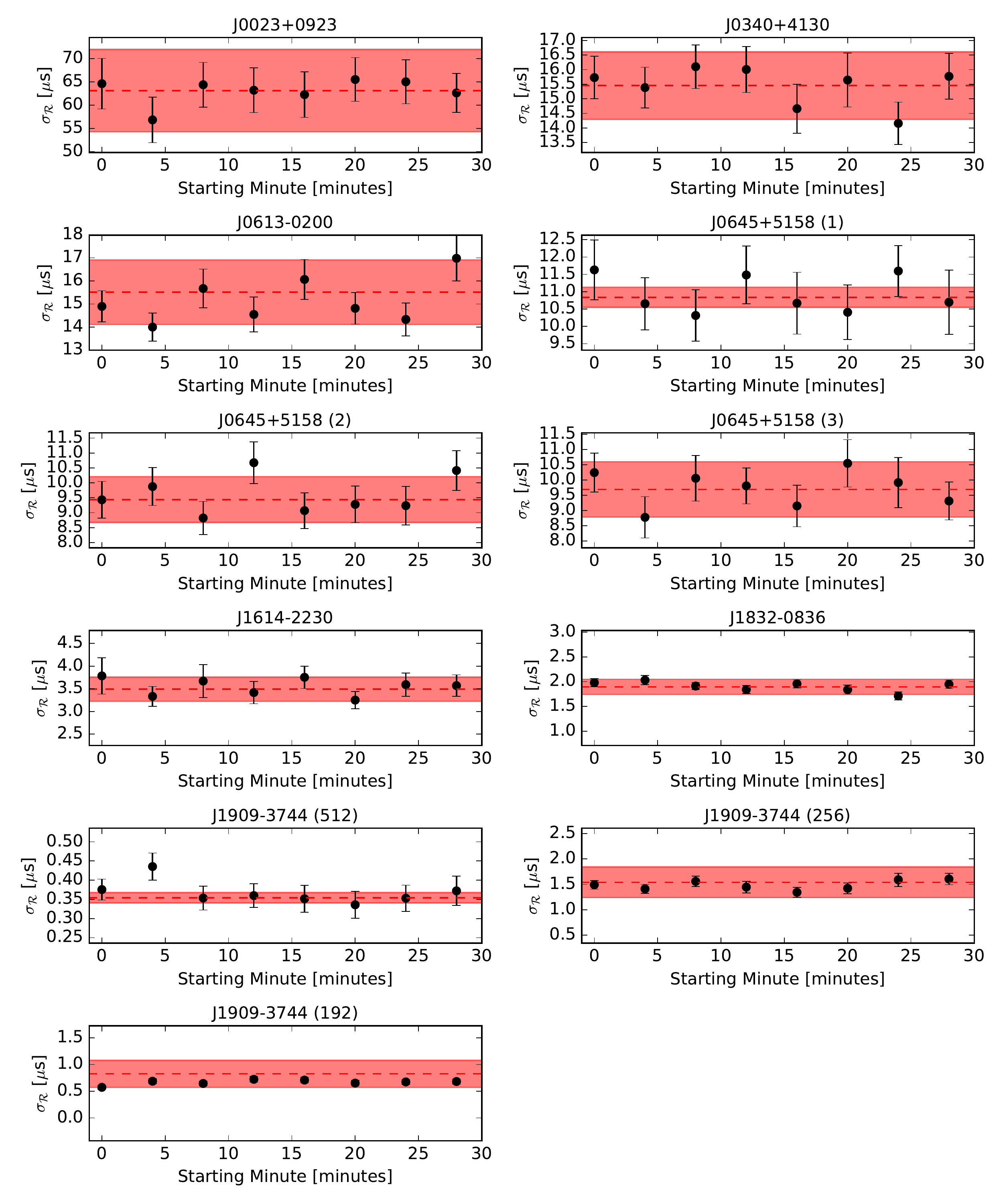}
    \caption{Comparison of \residerr for non-contiguous sets of pulsar timing residuals to contiguous set of timing residuals of equivalent length for each observation of each MSP. The red dashed line is the mean rms of the timing residuals for all equivalent contiguous sets of timing residuals and the red shaded region shows the one sigma standard deviation of the same. Each black point is the rms of a set of four minute integrated TOAs over 64 frequency channels separated by at least 30 minutes. The errorbars come from the standard deviation of \residerr calculated from 10000 bootstrapped samples of the 30~minute separated TOAs. The rms of the residuals here is larger than is seen in the NANOGrav data sets due to the smaller integration time used here \citep{Arzoumanian2015,Arzoumanian2018}. In each case there appears to be little difference between taking residuals from a contiguous observation and from non-contiguous observations. Only one point does not fall within the 1$\sigma$ error for the 512~frequency channel observation of PSR~J1909$-$3744. We do not know exactly why this is, but as this point is still within 2$\sigma$ of the average expected rms and varying RFI and ISM characteristics could lead to deviations.}
    \label{Bootstrapped_Resids_All}
\end{figure*}

For all MSPs in our sample we found that there was little difference in the rms of the timing residuals, \residerr, when using contiguous TOAs when compared to non-contiguous TOAs as expected. Errors on the rms of each set of residuals comes from the standard deviation of the distribution of rms residuals. All sets of non-contiguous timing residuals match within $1\sigma$ of the expected \residerr for our choice of 64 frequency channels and four minutes integrations, with the exception of one set of timing residuals from the full bandwidth observation of PSR~J1909$-$3744, as shown in Figure \ref{Bootstrapped_Resids_All}. The rms of the residuals here is larger than is seen in the NANOGrav data sets due to the smaller integration time used here \citep{Arzoumanian2015,Arzoumanian2018}.

It is possible that this particular set of non-contiguous residuals suffers from blower than average S/N compared to other sets of residuals, or from small amounts of RFI contamination, leading to larger errors on the timing residuals for this set, although there is nothing obvious for this set that shows this. As this point is still within $2\sigma$ of the expected \residerr, and all other points are with $1\sigma$, this point shows only that it is possible to get unlucky scintillation or RFI during observations.

This result is promising as it suggests that time-continuity of TOAs has a small, if any, effect on precision pulsar timing. While the gaps between CHIME TOAs will be much larger than those in this study, it suggests that they may not significantly affect the achievable timing precision. Additionally, if we had found DM variations on hour-long timescales it could pose problems for CHIME to track daily DM changes. There are of course many other considerations such as pulse profile evolution and observing frequency that will have to be considered along with these effects.

Our results do show that if sections of the observation need to be dropped due to RFI contamination, the timing residual precision will not be affected beyond the expected \residerr~$\propto 1/\sqrt{N}$. Additionally, if observations where a pulsar were only observed for short periods of time to try to observe only when it is scintillating brightly, the timing residual precision would not be affected due to the on-off nature of this observation, and would be improved due to higher S/N of brightly scintillated pulses.

\section{Conclusions} \label{conclusion}

We have examined various noise parameters commonly seen in pulsar timing residuals, determined scintillation parameters, some for the first time, looked for DM variations on hour-long timescales, and analyzed the impact of non-contiguous timing residuals using a unique set of multi-hour continuous observations of seven different MSPs. The major conclusion from our analyses are summarized below.
\begin{itemize}
    \item We present new measurements of \scintnu for all MSPs in our sample, some for the first time, as well as measurements and lower limits of \scintt for all MSPs for the first time. We find that 2-D Gaussian fitting gives more robust scintillation parameters than 1-D Gaussian fitting. We also find that the scintle drift rates for PSRs~J0023$+$0923 and J1614$-$2230 suggests that the line of sight is not well modeled by a uniform medium, and for J1614$-$2230 a single scattering screen is not sufficient to describe the line of sight. Additionally we are able to report values of \DISSerr in a regime where we are not dominated by the finite scintle effect for some of the MSPs in our sample.
    \item We estimate \jittererr of all MSPs using two different methods and find that the maximum likelihood method yields a more constraining result. Additionally we are limited by the low S/N of our observations, showing the importance of refractive scintillation on estimating pulse jitter at higher frequencies.
    \item We find the DM measured for each MSP in our sample does not vary within our sensitivity limits on timescales of hours, as expected.
    \item There is little difference in \residerr of timing residuals that are non-contiguous in time when compared with equivalent timing residuals that are contiguous. Our result is promising for instruments like CHIME restricted to short but frequent observations and also show that should a section of an observation be removed, the timing precision of the residuals will not be significantly affected.
    \item Given the results, particularly of the pulse jitter and DM variations on these long observations, we find that the TOA variations in these long data sets are consistent with the assumed breakdown into template-fitting error, jitter error, and DISS error, despite the limiting S/N of the data set.
\end{itemize}

The nature of a continuous long observation of a single pulsar, particular an MSP, allows us to study a wide variety of noise parameters commonly seen in pulsar timing as well properties of the ISM that are often difficult to otherwise probe. Additionally, continuous long observations are the only way to measure the scintillation timescale of many pulsars, an important characteristic for determining white noise caused by scintillation in pulsar timing residuals, and necessary to quantify for precision pulsar timing. In order to better constrain the noise parameters of MSPs as well as probe the ISM, it will be necessary to perform continuous long observations of many MSPs at multiple wavelengths.

\section*{Acknowledgements} \label{acknowledgements}
We would like to thank David Nice for valuable comments and suggestions. We would also like to thank the referee for useful comments that helped to improve the text. This work was supported by NSF Award OIA-1458952. J.M.C., M.T.L., M.A.M., B.J.S., and J.K.S are members of the NANOGrav Physics Frontiers Center which is supported by NSF award 1430284. B.J.S. acknowledges support from West Virginia
University through the STEM Mountains of Excellence Fellowship. The National Radio Astronomy Observatory and Green Bank Observatory are facilities of the National Science Foundation operated under cooperative agreement by Associated Universities, Inc. 

\section*{Software} \label{software}
{\textit{Software}: PSRCHIVE \citep{Hotan2004, VanStraten2012}, PyPulse \citep{LamPyPulse}, Scipy \citep{Jones2001}, Matplotlib \citep{Hunter2007}, TEMPO, TEMPO2 \citep{tempo2}}


\bibliography{MSP_Long_Obs_V12}{}

\appendix

\section{Binary Orbital Parameters}\label{BinaryErrs}

While our formalism for the short timescale timing model as described in \S \ref{whitenoise} is the same as is used in \cite{Lam2016a}, we have assumed that the fit to the timing residuals is well described by a quadratic as shown in Eq. \ref{shorttermTOAs}. However, for MSPs in a binary system, the observed pulse period can be Doppler shifted by some amount \citep{Handbook},
\begin{equation} \label{dopplercor}
\begin{split}
    \sigma_{P_{b}} \sim P \frac{\delta \nu_{\parallel}}{c} \sim \frac{2 \pi P}{c} \delta \left ( \frac{a~{\rm{sin}}~i}{P_{b}} \right) \sim \frac{2 \pi P a~{\rm{sin}}~i}{c P_{b}} \sqrt{\left (\frac{\delta a}{a} \right)^{2} + \left (\frac{\delta {\rm{sin}}~i}{{\rm{sin}}~i} \right)^{2} + \left (\frac{\delta P_{b}}{P_{b}} \right)^{2}} \\ \sim 72.7~{\rm{ns}}~P_{\rm{ms}}~a_{\rm{lsec}}~{\rm{sin}}~i~P_{b,{\rm{day}}}^{-1} \sqrt{\left (\frac{\delta a}{a} \right)^{2} + \left (\frac{\delta {\rm{sin}}~i}{{\rm{sin}}~i} \right)^{2} + \left (\frac{\delta P_{b}}{P_{b}} \right)^{2}}.
\end{split}
\end{equation}
Here $a$ is the semimajor axis, $i$ is the inclination angle, and $P_{b}$ is the binary orbital period, and we assume that the errors on the binary parameters are uncorrelated. 

However, the error induced by this in the timing residuals that are fit for in Eq. \ref{intrinsicTOAs} will follow a cubic of $\sim \sigma_{P_{b}} (T/P_{b})^{3}$, where $T$ is the length of the observation. Due to the length of the observations used in this analysis, $(T/P_{b})$ may be quite large. In fact, for PSR~J0023$+$0923, the MSP with the shortest binary period in this work, $\sim 200$~ minutes, $(T/P_{b}) = 0.855$. However, for PSR~J0023$+$0923, $\sigma_{P_{b}} \approx 1 \times 10^{-4}$~ns, so the total error is $\ll 1$~ns. Out of all pulsars in this work, the largest binary parameter error is for PSR~J0613$-$0200 of 0.08~ns. As this is much less than \residerr for all of the observations in this work, the error induced by the binary orbit parameters is negligible.

\end{document}